\documentclass[preprint2]{aastex}

\shorttitle{Andromeda~XXVII}
\shortauthors{Cusano et al.}
\usepackage{color}

\begin{document}

\title{VARIABLE STARS AND STELLAR POPULATIONS IN ANDROMEDA~XXVII: IV. AN OFF-CENTERED, DISRUPTED GALAXY  \altaffilmark{*}}

\author{ FELICE CUSANO\altaffilmark{1}, ALESSIA GAROFALO\altaffilmark{1,2}, GISELLA CLEMENTINI\altaffilmark{1},
MICHELE CIGNONI\altaffilmark{3}, TATIANA MURAVEVA\altaffilmark{1}, GIANNI TESSICINI\altaffilmark{1}, VINCENZO TESTA\altaffilmark{4}, DIEGO PARIS\altaffilmark{4},
LUCIANA FEDERICI\altaffilmark{1}, 
MARCELLA MARCONI\altaffilmark{5}, VINCENZO RIPEPI\altaffilmark{5}, ILARIA MUSELLA\altaffilmark{5}}

\affil{$^1$INAF- Osservatorio Astronomico di Bologna, Via Gobetti 93/3, I - 40129 Bologna, Italy}
\email{felice.cusano@oabo.inaf.it}
\affil{$^2$ Dipartimento di Fisica e Astronomia, Universit\`a di Bologna,  viale Berti Pichat, 6/2, I - 40127 Bologna, Italy}
\affil{$^3$ Dipartimento di Fisica, Universit\`a di Pisa, Largo Bruno Pontecorvo, 3, 56127 Pisa PI}
\email{}
\affil{$^4$INAF- Osservatorio Astronomico di Roma, Via di Frascati 33
00040 Monte Porzio Catone, Italy}
\email{}
\affil{$^5$INAF- Osservatorio Astronomico di Capodimonte, Salita Moiariello 16, 
I - 80131 Napoli, Italy}
\email{}
\altaffiltext{*}{Based on data collected  with the Large Binocular Cameras at the Large Binocular Telescope, PI: G. Clementini}

\begin{abstract}
\noindent 
We present $B$ and $V$ time series photometry of the M31 satellite galaxy Andromeda~XXVII (And~XXVII)  that we observed with
the Large Binocular Cameras of the Large Binocular Telescope. 
In the field of And~XXVII we have discovered a total of 90 variables: 89 RR Lyrae stars and 1 Anomalous Cepheid.
The average period of the fundamental mode RR Lyrae stars (RRab)  $\langle$P$_{\rm ab}\rangle$=0.59 d ($\sigma$=0.05 d)
and the period-amplitude diagram place And~XXVII in the class of Oosterhoff I/Intermediate objects.
Combining information from the color-magnitude diagram (CMD) and the variable stars we find evidence
for a single old and metal poor stellar population with [Fe/H]$\sim -1.8$ dex and  t$\sim$13 Gyr in And~XXVII. 
The spatial distribution of RR Lyrae and red giant branch (RGB) stars gives clear indication that And~XXVII
is a completely disrupted system. This is also supported by 
the  spread  observed along the line of sight  in the distance to the RR Lyrae stars.
The highest concentration of RGB and RR Lyrae stars is found in a circular area of 4 arcmin in radius, 
centered about 0.2 degrees in south-east direction from  \citet{rich11}  center coordinates of And~XXVII. 
The CMD of this region is well defined with a prominent RGB and 15 RR Lyrae stars (out of the 18 found in the region) tracing a 
very tight horizontal branch at  $\langle V(RR) \rangle$ = 25.24 mag~~ $\sigma$=
0.06 mag (average over 15 stars). 
We show that And XXVII well proposes as a candidate building block of the M31 halo.

\end{abstract}

\keywords{galaxies: dwarf, Local Group 
---galaxies: individual (Andromeda~XXVII)
---stars: distances
---stars: variables: other
---techniques: photometric}

\section{INTRODUCTION}\label{sec:intro}
In the $\Lambda$-cold dark matter (CDM) scenario, 
galaxies are formed by  hierarchical assembling of smaller structures
\citep[e.g.][]{bul05, ann2016,stier2017}.
The dwarf satellites of the Andromeda galaxy (M31) can
help to constrain the origin and the fate of M31.
Through the characterization of the resolved stellar populations 
and the variable stars in these systems it 
is possible  to trace the  global context of merging and accretion episodes occurred and still
occurring in the  M31 environment \citep{mart2013}.

This is the fourth  paper in our series on the M31 satellites based on 
$B$ and $V$ time-series photometry obtained with the Large Binocular Cameras (LBC) of  the Large Binocular Telescope (LBT). 
Details on the survey and results from the study of  Andromeda~XIX (And~XIX), Andromeda~XXI (And~XXI) and Andromeda~XXV (And~XXV)
were presented in \citet[][Paper~I]{cus2013}, \citet[][Paper~II]{cus2015} and \citet[][Paper~III]{cus2016}, respectively. In this paper we  
report  results on the dwarf spheroidal galaxy (dSph) Andromeda~XXVII (And~XXVII),
that was discovered by \citet{rich11} in the context of the PAndAS survey. The galaxy is located near 
a portion of Andromeda's North-West stream \citep[NW; see Fig.~1 of ][]{rich11}. 
These authors claimed that And~XXVII is in the process of being tidally disrupted by M31 and derived 
only a lower limit of $\ge757\pm45$ kpc for its heliocentric distance due to the difficulty in measuring the magnitude of the galaxy horizontal branch (HB).
\citet{iba2013} later  found that And~XXVII is  a member of the thin plane of satellites identified in M31 
\citep[Great Plane of Andromeda, GPoA, following the definition of][]{paw2013}.
The heliocentric distance of And~XXVII was revised using  tip of the Red Giant Branch (RGB) stars by \citet{con12} who 
estimated a value of $1255^{+42}_{-474}$ kpc that
would place the galaxy outside the M31 complex. The RGB tip of  And~XXVII is very scarcely populated likely due 
to the disrupting nature of the galaxy. This may have significantly affected \citet{con12} determination of distance.
\citet{col13} measured a metallicity for  And~XXVII of [Fe/H]$=-2.1\pm0.5$ dex estimated from the 
Calcium triplet (CaII) in eleven probable member stars.  
The radial velocity of And~XXVII measured by the same authors  is:  v$_r$= $-$539.6 km s$^{-1}$  with $\sigma=14.8$ km s$^{-1}$.
 \citet{col13}  warned that given the disrupting nature of the object, their 
estimate is very uncertain. They also identified a significant kinematic substructure
around $\sim-500$ km s$^{-1}$.
Very recently \citet{mart2016} from a re-analysis of the PAndAS data stated that And~XXVII 
is likely  a system that is in the final phase of tidal disruption. The wide-field LBT observations of And~XXVII presented in this paper lend further 
support to this evidence from the analysis of both color magnitude diagram (CMD) and  spatial 
distribution of the variable stars.

The paper is organized as follows: Section~2 describes the data collection and processing, Section~3 describes the identification of the variable stars and their characterization.
The galaxy CMD is presented in Section~4. Distance and galaxy structure are discussed in Sections~5 and 6. Conclusions are provided in Section~7. 

\begin{table*}
\begin{center}
\caption[]{Log of And~XXVII  observations}
\label{t:obs}
\begin{tabular}{l c c c c }
\hline
\hline
\noalign{\smallskip}
   Dates                 & {\rm Filter}  & N   & Exposure time &  {\rm Seeing (FWHM)}    \\
 	                 &		 &     & (s)           &    {\rm (arcsec)}\\
\noalign{\smallskip}
\hline
\noalign{\smallskip}
  October 20-24, 2011  &   $B$      & 87     & 400    & 0.8-1  \\ 
  November 27-28, 2011 &   $B$      &    5    &   400     & 0.8-1 \\
  
                       &             &   &        &        \\  
  October 20-24, 2011  &   $V$      &    86   & 400    & 0.8-1  \\ 
  November 27-28, 2011 &   $V$      &    4    &   400     & 0.8-1 \\
  
\hline
\end{tabular}
\end{center}
\normalsize
\end{table*}

\section{OBSERVATIONS AND DATA REDUCTION}
Time series observations in the $B$ and $V$ bands of the field around the center coordinates 
of And~XXVII \citep[R.A.=$00^{\rm h}37^{\rm m}27^{\rm s}$, decl.=+45$^{\circ}23{\arcmin}13{\arcsec}$; J2000, ][]{rich11}  
were carried out  in October-November 2011 (see Table~\ref{t:obs} for the complete log)
at the LBT equipped with the LBC. 
The total LBC's field of view (FoV) covers an area of $\sim 23\arcmin\times23\arcmin$.
Observations were obtained under good sky and seeing conditions (see Table~\ref{t:obs}).
The blue camera  (LBC-B) was used to acquire the $B$ images, while $V$  imaging was obtained with the red camera  (LBC-R).
Bias, flat fielding and distortion corrections of the raw frames were performed with a dedicated pipeline developed at
INAF-OAR\footnote{$http://lbc.oa-roma.inaf.it/commissioning/index.html$}.
We performed PSF photometry of the pre-processed images using the  \texttt{DAOPHOT - ALLSTAR - ALLFRAME} packages \citep{ste87,ste94} as described
in Paper~I. Photometric calibration was performed using the Landolt standard fields L92 and SA113,  observed during the same observing run. 
The calibration equations\footnote{
$B-b=27.696-0.113\times(b-v)$ r.m.s=0.03,\\
$V-v=27.542-0.060\times(b-v)$ r.m.s=0.03}
are consistent with those
derived in Paper~I, once differences in air-mass and exposure times are accounted for.

\section{VARIABLE STARS}
\begin{figure}[t!]
\centering
\includegraphics[trim=0 0 0 0 clip, width=1.\linewidth]{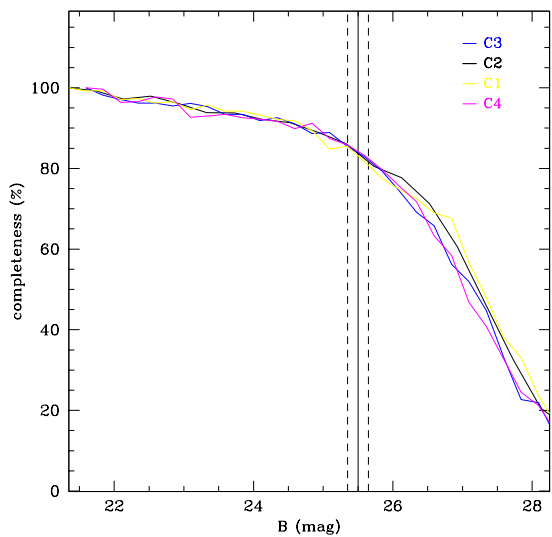}
\caption{ Completeness of the  photometric catalogs  in the four CCDs (hereinafter C1, C2, C3 and C4)
of the LBC mosaic. The blue line is  C3, the yellow line is C1, the magenta line is C4 and 
the black line is C2. The vertical solid line marks the average $B$ magnitude of the RR Lyrae stars, dashed lines are the 1 $\sigma$ boundaries.}
\label{fig:compl}
\end{figure}

Variable stars were identified in our photometric catalogs using the same procedure 
as described in Paper~I. We used the  variability  index computed in DAOMASTER \citep{ste94}
to search for variable sources. 
 We selected as candidate variables, stars  with a variability index (VI) $>$ 1.4, 
based on our previous experience with  other M31 satellites (And~XIX, And~XXI, And~XXV)
and LBT photometry.
The average value  for non variable 
stellar objects placed on the horizontal branch (HB) and satisfying the  quality parameters, $\chi \le  1.5$,  $-0.35 < {\rm Sharpness} < 0.35$ (see Section~\ref{sec:cmd}),  
is $\langle {\rm VI} \rangle= 1.03$, $\sigma=0.12$.
The value of  VI  used to select candidate variable stars is thus 3 $\sigma$ above the average 
value for HB non variable stars. The selection of variable stars was performed  on the  photometric
catalog for objects satisfying to the quality parameters ($\chi$ and  Sharpness) mentioned above. The total number of candidate variables that passed our 
selection criteria is 754. 
The $B$ and $V$ light curves of the candidate variables
were then analyzed using the Graphical  Analyzer  of  Time Series  package  (GRaTIS),  custom  software  developed  at  the
Bologna Observatory by P. Montegriffo \citep[see, e.g.,][]{clm00}.
From this analysis, 90 sources were confirmed to vary. Of them  89 are classified as RR Lyrae stars and one as Anomalous Cepheid (AC),
 based on the pulsation properties and the position in the CMD.
 The lowest value of VI for
a bona fide RR Lyrae stars identified in this work is VI=1.52, 
hence the VI limit  we chose to select candidate variable stars is  loose
enough to ensure we did not miss candidate variable. Therefore,
the completeness of our RR Lyrae stars catalog is  mostly driven 
by the photometric  completeness at the HB level. 
This was estimated using artificial star test and 
at the level of HB is $\sim83\%$ as shown in Figure~\ref{fig:compl}. The Figure shows  the trend of the completeness 
in the  $B$-band magnitude for each of the four CCDs of the LBC mosaic, clearly  
there is no significant difference  among the four CCDs especially at the average magnitude of the RR Lyrae stars (solid vertical line). 
Finally the incompleteness of our RR Lyrae stars catalog is of  $\sim 17\%$ 
 hence we are probably missing from 10 to 20  RR Lyrae stars in And~XXVII field.

\begin{figure*}
\centering
\includegraphics[trim=0 0 0 0, width=1.\linewidth]{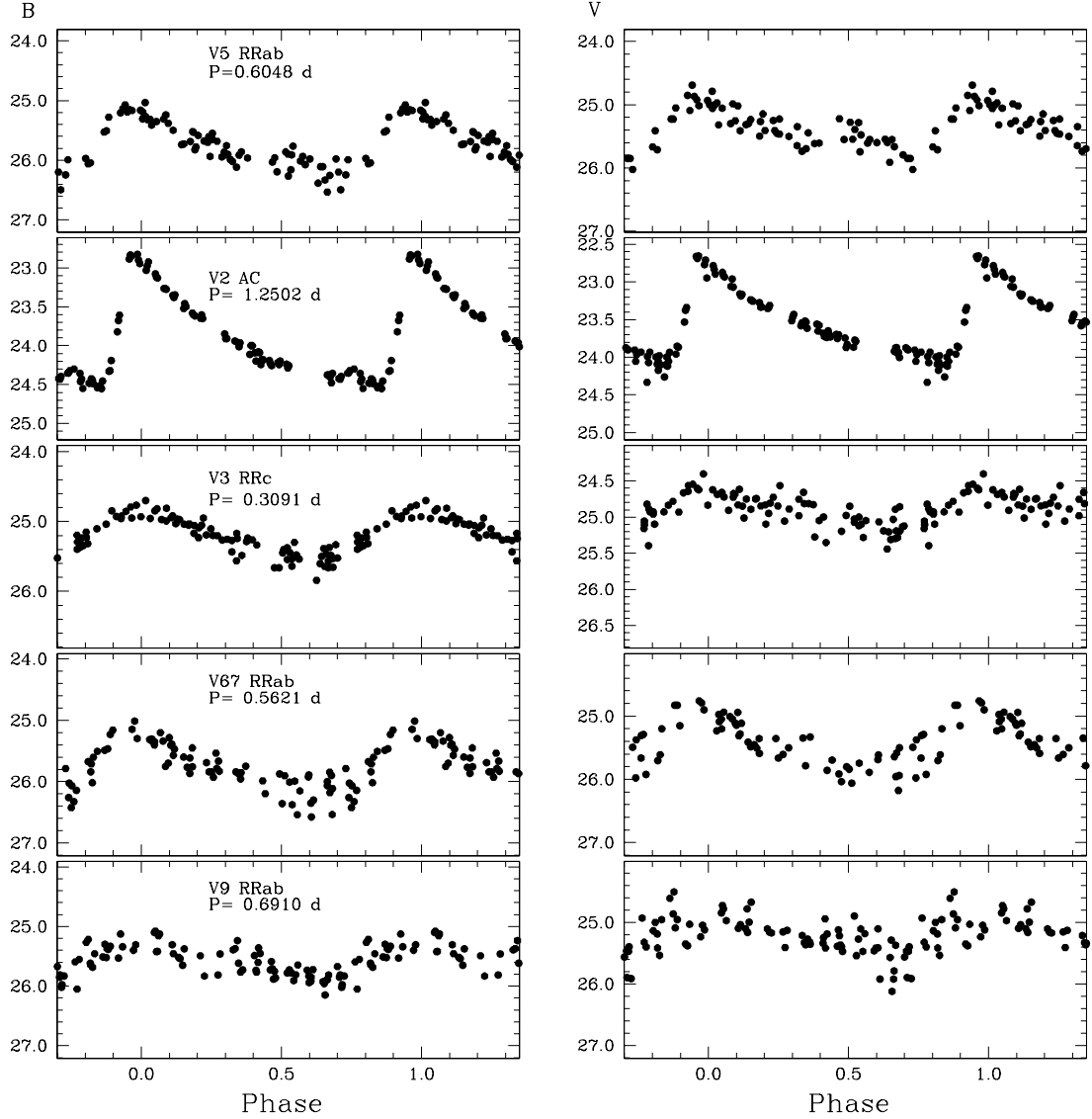}
\caption{Examples of $B$- (left panels) and $V$-band (right panels)  light curves for different types of variable stars identified 
in And~XXVII (from top to bottom: fundamental-mode RR Lyrae star, 
anomalous Cepheid, first overtone RR Lyrae star, V67 the only fundamental-mode RR Lyrae star with  $D_m<3$ and V9 the RRab with the highest $D_m$ value). 
Typical internal errors of the single-epoch data range from 0.01 at $B\sim$  22.8 mag (corresponding to the maximum light of the AC) to 0.30 mag at $B\sim$  26.8 mag 
(corresponding to the minimum light of a fundamental mode RR Lyrae), and similarly,  from 0.01 mag at $V\sim$  22.6 mag  to 0.30 mag at $V\sim$  26.8 mag.}
\label{fig:lca-examples}
\end{figure*}

Coordinates and properties of the confirmed variable stars are listed in Table~\ref{t:1} where  
column 1  gives the star identifier; columns 2 and 3 are, respectively, the right ascension and declination (J2000.0 epoch)  
obtained from our astrometrized catalogs; column 4 provides the type of variability with a  question mark identifying stars whose classification is uncertain; 
columns 5 and 6 give, respectively, the period and the Heliocentric Julian Day (HJD) of maximum light; 
columns 7 and 9 are the intensity-averaged mean $B$ and $V$ magnitudes; columns 8 and 10 list the corresponding amplitudes of  light variation.
Examples of light curves, in both filters, are shown in  Figure~\ref{fig:lca-examples}.
 Observations have been performed in five consecutive nights in October and two consecutive nights in November.
The data are mostly biased by the one day alias due to the alternation of night-day  observations and are thus marginally affecting
the period determination of the RR Lyrae stars that in all cases is $<1$ day.
The full catalog of light curves is available in the  electronic version of the paper.

\begin{table*}
\caption[]{Identification and properties of the variable stars detected in And~XXVII.}
\tiny
\label{t:1}
\begin{center}
\begin{tabular}{l c c l l c c c c c }
\hline
\hline
\noalign{\smallskip}
 Name & $\alpha$            &$\delta$       & Type & ~~~P        & Epoch (max)      & $\langle B \rangle$ &  A$_{B}$ & $\langle V \rangle$  & A$_{V}$ \\ 
     &	(2000)              & (2000)        &	   &~(days)      & HJD ($-$2455000) & (mag)               & (mag)               & (mag)   &  (mag)     \\
\noalign{\smallskip}
\hline
\noalign{\smallskip}
   V1  & 00:37:31.893  &  +45:23:54.12 &  RRab &  0.6171  & 53.792   &     25.38   & 0.62  & 24.94   &  0.49	\\
   V2  & 00:37:32.950  &  +45:22:58.27 &  AC  &   1.2502  & 53.085   &     23.79   & 1.73  & 23.48   &  1.43	\\
   V3  & 00:37:33.655  &  +45:24:40.61 &  RRc &   0.3091  & 53.633   &     25.19   & 0.74  & 24.91   &  0.53	\\
   V4  & 00:37:19.667  &  +45:24:27.33 &  RRc &   0.3857  & 53.272   &     25.38   & 0.82  & 25.08   &  0.59	\\
   V5  & 00:37:36.643  &  +45:22:58.03 &  RRab &  0.6048  & 54.080   &     25.72   & 1.05  & 25.37   &  0.82	\\
   V6  & 00:37:19.271  &  +45:21:17.88 &  RRc &   0.3642  & 53.600   &     25.61   & 0.59  & 25.32   &  0.46	\\
   V7  & 00:37:28.721  &  +45:26:12.74 &  RRab &  0.5798  & 53.762   &     25.78   & 1.04  & 25.25   &  0.55	\\
   V8  & 00:37:25.769  &  +45:19:50.86 &  RRab &  0.5999  & 53.421   &     25.55   & 0.94  & 25.14   &  0.62	\\
   V9  & 00:37:22.045  &  +45:26:39.96 &  RRab &  0.6910  & 53.292   &     25.56   & 0.66  & 25.19   &  0.66	\\
   V10 & 00:37:23.724  &  +45:19:30.90 &   RRab &  0.6226  & 53.817   &    25.74   & 0.52  & 25.33   &  0.37	\\
   V11 & 00:37:16.889  &  +45:26:08.05 &   RRc &   0.3552  & 53.428   &    25.41   & 0.69  & 25.06   &  0.42	\\
   V12 & 00:37:19.295  &  +45:19:31.97 &   RRc &   0.2432  & 53.856   &    25.30   & 0.69  & 25.05   &  0.57	\\
   V13 & 00:37:37.243  &  +45:27:03.68 &   RRc &   0.3635  & 53.735   &    25.38   & 0.72  & 24.99   &  0.53	\\
   V14 & 00:37:38.799  &  +45:26:48.76 &   RRab &  0.6710  & 53.271   &    25.22   & 0.94  & 24.76   &  0.70	\\
   V15 & 00:37:44.006  &  +45:21:10.15 &   RRab &  0.5645  & 53.783   &    25.78   & 0.97  & 25.53   &  0.85	\\
   V16 & 00:37:21.577  &  +45:18:40.74 &   RRab &  0.5881  & 53.459   &    25.72   & 1.16  & 25.35   &  0.78	\\
   V17 & 00:37:46.582  &  +45:23:51.84 &   RRab &  0.6586  & 53.390   &    25.64   & 0.67  & 25.30   &  0.59	\\
   V18 & 00:37:29.435  &  +45:28:15.86 &   RRab &  0.5793  & 53.425   &    25.57   & 1.09  & 25.21   &  0.89	\\
   V19 & 00:37:30.249  &  +45:18:03.82 &   RRab &  0.6181  & 53.987   &    25.33   & 1.30  & 25.05   &  1.18	\\
   V20 & 00:37:07.677  &  +45:26:04.87 &   RRab &  0.5612  & 53.679   &    25.30   & 1.49  & 24.92   &  0.99	\\
   V21 & 00:37:40.518  &  +45:27:58.79 &   RRc &   0.4357  & 53.691   &    25.42   & 0.46  & 24.99   &  0.31	\\
   V22 & 00:37:12.966  &  +45:18:22.55 &   RRc &   0.2925  & 53.985   &    25.33   & 0.67  & 25.12   &  0.57	\\
   V23 & 00:37:38.962  &  +45:28:28.95 &   RRab &  0.5570  & 53.857   &    25.11   & 1.14  & 24.70   &  0.96	\\
   V24 & 00:37:06.459  &  +45:19:45.37 &   RRab &  0.5746  & 53.559   &    25.41   & 1.19  & 25.10   &  0.96	\\
   V25 & 00:37:45.060  &  +45:18:43.87 &   RRab &  0.7150  & 53.391   &    25.52   & 0.71  & 25.11   &  0.70	\\
   V26 & 00:37:01.535  &  +45:24:21.95 &   RRab &  0.5535  & 53.588   &    25.66   & 1.00  & 25.26   &  1.03	\\
   V27 & 00:37:45.003  &  +45:28:36.34 &   RRc &   0.3938  & 53.147   &    25.52   & 0.56  & 25.10   &  0.35	\\
   V28 & 00:37:52.395  &  +45:19:57.61 &   RRab &  0.5870  & 53.910   &    25.58   & 0.95  & 25.13   &  0.87	\\
   V29 & 00:37:17.317  &  +45:30:03.48 &   RRc &   0.3601  & 54.076   &    25.34   & 0.44  & 24.88   &  0.38	\\
   V30 & 00:36:59.737  &  +45:25:39.13 &   RRab &  0.5541  & 53.887   &    25.50   & 1.04  & 25.14   &  0.69	\\
   V31 & 00:37:56.368  &  +45:25:06.03 &   RRab &  0.5408  & 53.803   &    25.56   & 1.05  & 25.25   &  0.98	\\
   V32 & 00:37:57.393  &  +45:21:23.21 &   RRc? &  0.2793  & 53.931   &    25.44   & 0.47  & 25.27   &  0.47	\\
   V33 & 00:37:56.332  &  +45:20:25.79 &   RRab &  0.5492  & 53.824   &    25.50   & 0.82  & 25.23   &  0.50	\\
   V34 & 00:37:00.936  &  +45:18:26.79 &   RRab &  0.5579  & 53.875   &    25.54   & 1.46  & 25.18   &  1.00	\\
   V35 & 00:37:58.868  &  +45:20:16.64 &   RRc &   0.3579  & 53.544   &    25.67   & 0.58  & 25.30   &  0.43	\\
   V36 & 00:38:00.070  &  +45:21:13.33 &   RRc &   0.3641  & 53.781   &    25.60   & 0.66  & 25.21   &  0.67	\\
   V37 & 00:38:00.481  &  +45:24:46.02 &   RRab &  0.5996  & 53.525   &    25.35   & 1.31  & 24.99   &  0.96	\\
   V38 & 00:37:52.645  &  +45:17:32.73 &   RRc &   0.3987  & 53.928   &    25.27   & 0.63  & 24.92   &  0.45	\\
   V39 & 00:36:55.883  &  +45:27:09.36 &   RRc &   0.3534  & 53.504   &    25.51   & 0.82  & 25.14   &  0.56	\\
   V40 & 00:38:02.723  &  +45:24:46.06 &   RRc &   0.3395  & 53.616   &    25.50   & 0.72  & 25.28   &  0.59	\\
   V41 & 00:37:13.142  &  +45:14:50.52 &   RRab &  0.5888  & 53.486   &    25.50   & 1.05  & 25.20   &  1.02	\\
   V42 & 00:36:51.394  &  +45:25:50.83 &   RRc &   0.3550  & 53.660   &    25.55   & 0.57  & 25.27   &  0.56	\\
   V43 & 00:37:56.003  &  +45:29:09.95 &   RRab &  0.6087  & 53.686   &    25.66   & 0.97  & 25.24   &  0.72	\\
   V44 & 00:37:25.859  &  +45:32:50.55 &   RRab? & 0.6312  & 53.322   &    25.37   & 0.45  & 25.12   &  0.53	\\
   V45 & 00:36:49.108  &  +45:20:47.83 &   RRab &  0.5315  & 53.758   &    25.37   & 1.28  & 25.03   &  1.18	\\
   V46 & 00:37:24.064  &  +45:13:16.69 &   RRab &  0.5855  & 54.116   &    25.78   & 0.82  & 25.46   &  0.90	\\
   V47 & 00:38:02.920  &  +45:18:20.73 &   RRab &  0.5293  & 53.342   &    25.67   & 1.14  & 25.29   &  0.70	\\
   V48 & 00:38:06.624  &  +45:20:36.47 &   RRc  &  0.3501  & 53.816   &    25.58   & 0.54  & 25.26   &  0.45	\\
   V49 & 00:38:05.333  &  +45:19:20.00 &   RRc  &  0.3541  & 53.015   &    25.54   & 0.50  & 25.21   &  0.31	\\
   V50 & 00:36:48.952  &  +45:19:02.07 &   RRab &  0.6451  & 53.359   &    25.56   & 1.17  & 25.14   &  0.76	\\
   V51 & 00:37:13.308  &  +45:13:20.85 &   RRc &   0.3558  & 53.410   &    25.53   & 0.61  & 25.28   &  0.51	\\
   V52 & 00:36:47.422  &  +45:27:31.80 &   RRab &  0.5627  & 53.761   &    25.20   & 1.21  & 24.84   &  0.85	\\
   V53 & 00:36:52.857  &  +45:30:18.98 &   RRab &  0.5980  & 53.706   &    25.42   & 0.78  & 25.06   &  0.69	\\
   V54 & 00:38:11.223  &  +45:21:26.58 &   RRab &  0.5754  & 54.079   &    25.41   & 1.10  & 25.12   &  1.02	\\
   V55 & 00:37:09.934  &  +45:12:53.19 &   RRab &  0.6226  & 53.920   &    25.53   & 0.79  & 25.24   &  0.57	\\
   V56 & 00:38:02.164  &  +45:30:14.28 &   RRab &  0.5794  & 53.853   &    25.60   & 0.95  & 25.26   &  0.84	\\
   V57 & 00:36:42.634  &  +45:21:18.38 &   RRab &  0.5357  & 53.638   &    25.49   & 1.18  & 25.29   &  0.96	\\
   V58 & 00:38:10.002  &  +45:18:59.73 &   RRab &  0.5395  & 53.369   &    25.62   & 1.00  & 25.25   &  0.79	\\
   V59 & 00:37:43.102  &  +45:34:14.31 &   RRab &  0.6000  & 53.707   &    25.42   & 1.19  & 25.04   &  0.92	\\
   V60 & 00:36:39.192  &  +45:23:25.77 &   RRab &  0.5788  & 53.866   &    25.47   & 0.94  & 25.09   &  0.94	\\
   V61 & 00:38:16.025  &  +45:24:01.56 &   RRc? &  0.3830  & 53.841   &    25.43   & 0.75  & 25.20   &  0.76	\\
   V62 & 00:36:37.523  &  +45:23:46.74 &   RRc  &  0.4281  & 53.797   &    25.41   & 0.58  & 24.95   &  0.22	\\
   V63 & 00:38:15.170  &  +45:19:44.42 &   RRab &  0.6023  & 53.829   &    25.10   & 0.87  & 24.73   &  0.71	\\
   V64 & 00:36:42.559  &  +45:17:17.99 &   RRab &  0.5537  & 53.877   &    25.41   & 1.19  & 25.10   &  0.96	\\
   V65 & 00:38:07.107  &  +45:31:17.21 &   RRab &  0.6035  & 53.411   &    25.57   & 0.87  & 25.16   &  0.53	\\
   V66 & 00:38:17.987  &  +45:21:20.02 &   RRab &  0.7383  & 53.309   &    25.34   & 0.52  & 24.77   &  0.35	\\
   V67 & 00:38:14.697  &  +45:17:42.86 &   RRab &  0.5621  & 53.544   &    25.76   & 0.90  & 25.45   &  0.68	\\
   V68 & 00:36:45.329  &  +45:14:56.17 &   RRc  &  0.3682  & 53.304   &    25.49   & 0.67  & 25.21   &  0.50	\\
   V69 & 00:38:13.766  &  +45:16:28.40 &   RRab &  0.5715  & 53.933   &    25.70   & 0.90  & 25.33   &  0.98	\\
   V70 & 00:36:33.155  &  +45:24:24.17 &   RRc  &  0.3547  & 53.748   &    25.54   & 0.58  & 25.15   &  0.37	\\
   V71 & 00:38:20.719  &  +45:20:47.63 &   RRc  &  0.3594  & 53.929   &    25.45   & 0.66  & 25.15   &  0.58	\\
   V72 & 00:38:18.355  &  +45:27:54.69 &   RRab &  0.6236  & 53.679   &    25.64   & 0.84  & 25.25   &  0.71	\\
   V73 & 00:38:06.155  &  +45:32:45.88 &   RRab &  0.6524  & 53.969   &    25.57   & 1.22  & 25.27   &  1.16	\\
   V74 & 00:38:12.060  &  +45:15:24.61 &   RRab &  0.5616  & 53.699   &    25.56   & 0.96  & 25.12   &  0.56	\\
   V75 & 00:38:05.872  &  +45:13:29.96 &   RRc  &  0.3487  & 53.877   &    25.50   & 0.64  & 25.20   &  0.56	\\
   V76 & 00:36:35.681  &  +45:28:17.18 &   RRc  &  0.3642  & 53.833   &    25.53   & 0.65  & 25.20   &  0.48	\\
   V77 & 00:38:23.292  &  +45:22:27.00 &   RRab &  0.5306  & 53.267   &    25.65   & 1.15  & 25.30   &  0.96	\\
   V78 & 00:38:14.075  &  +45:15:18.82 &   RRc  &  0.3333  & 53.837   &    25.50   & 0.68  & 25.24   &  0.65	\\
   V79 & 00:36:29.952  &  +45:23:43.59 &   RRc? &  0.2924  & 53.494   &    25.37   & 0.70  & 25.19   &  0.48	\\
   V80 & 00:38:23.752  &  +45:20:48.11 &   RRc  &  0.3648  & 53.352   &    25.61   & 0.74  & 25.31   &  0.64	\\
   V81 & 00:36:30.046  &  +45:25:46.60 &   RRab &  0.5595  & 53.968   &    25.26   & 1.01  & 24.89   &  0.82	\\
   V82 & 00:38:09.009  &  +45:12:59.08 &   RRc  &  0.2547  & 53.786   &    25.49   & 0.69  & 25.24   &  0.38	\\
   V83 & 00:38:27.464  &  +45:23:03.51 &   RRab &  0.5628  & 53.780   &    25.50   & 1.15  & 25.16   &  0.92	\\
   V84 & 00:36:31.312  &  +45:28:59.64 &   RRab &  0.6628  & 53.385   &    25.33   & 1.00  & 24.83   &  0.71	\\
   V85 & 00:38:20.435  &  +45:15:59.67 &   RRab &  0.5575  & 53.929   &    25.87   & 1.38  & 25.48   &  1.11	\\
   V86 & 00:38:28.400  &  +45:20:51.17 &   RRab &  0.5453  & 54.120   &    25.73   & 1.10  & 25.29   &  0.99	\\
   V87 & 00:36:28.323  &  +45:15:54.01 &   RRab &  0.7117  & 53.641   &    25.45   & 0.77  & 25.10   &  0.56	\\
   V88 & 00:38:32.760  &  +45:26:30.92 &   RRab &  0.6063  & 53.786   &    25.47   & 0.73  & 25.07   &  0.56	\\
   V89 & 00:36:19.903  &  +45:21:49.53 &   RRab &  0.5905  & 53.809   &    25.29   & 1.68  & 25.01   &  1.47	\\
   V90 & 00:36:21.505  &  +45:18:27.19 &   RRab &  0.5959  & 53.657   &    25.41   & 1.03  & 25.15   &  0.64	\\
\hline 
\end{tabular}
\end{center}
\normalsize
\end{table*}

\subsection{RR LYRAE STARS}\label{sec:rrli}
\begin{figure}[t!]
\centering
\includegraphics[trim=40 15 0 0 clip, width=1\linewidth]{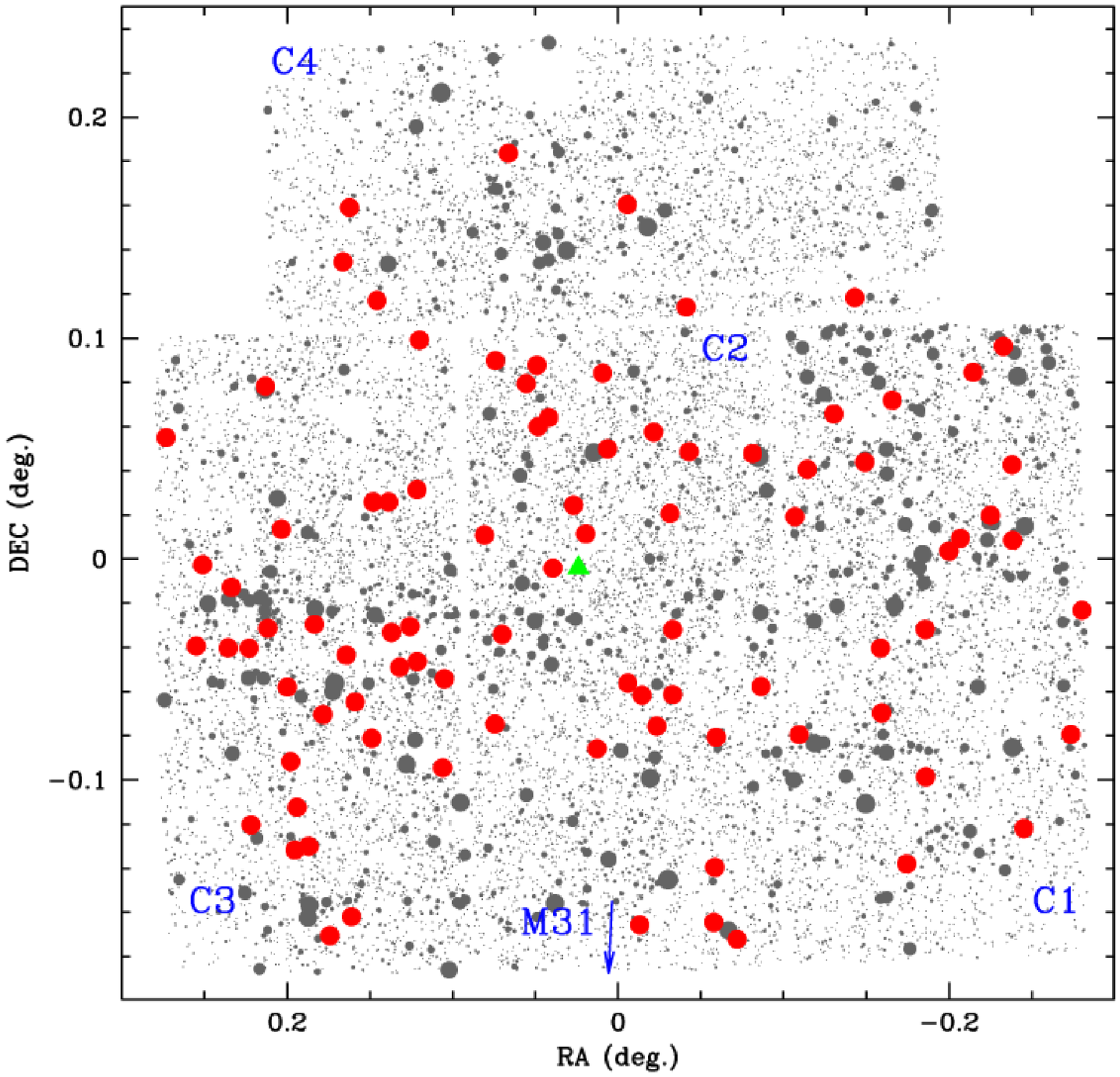}
\caption[]{Spatial distribution of the variable stars discovered in the whole LBT FoV. The 4 CCDs of the LBC mosaic are labeled.
Red filled circles mark the RR Lyrae stars, the green filled triangle is the AC.
Gray points are non variable sources, selected as to have DAOPHOT quality image parameters: $-0.35 \leq$ Sharpness $\leq$0.35 and $\chi < $1.5. 
Their size  is  inversely proportional to the  source $V$-magnitude.
Contamination by MW sources and background galaxies has not been removed (see Section~\ref{sec:cmd}).
 There are 32 RR Lyrae stars displayed in C3, 
 28 in C2,  21 in C1 and  7 in C4.
 The blue arrow points to the  direction of M31. North is up, east to the left. 
 
}
\label{fig:spa}
\end{figure}

In  the field  
 of And~XXVII  we have discovered a total of 
89 RR Lyrae stars, of which 58 are fundamental mode (RRab) and 31 are first-overtone (RRc) pulsators
according to  period and amplitude of light variation.
RR Lyrae stars are excellent tracers of an old  stellar population \citep[e.g.][and references therein]{cleme2010, marc2015}. Their presence in And~XXVII shows 
that this galaxy started forming stars $\geq$ 10 Gyr ago.
The spatial distribution of the variable stars discovered in the FoV of And~XXVII is shown if Figure~\ref{fig:spa}.
The highest concentration of RR Lyrae stars occurs in
CCD3 (labeled C3 in the figure) where there are 32 RR Lyrae to compare with 28 in C2, 21 in C1 and 7 in C4.
Furthermore, the very sparse spatial distribution of these variable stars  
 suggests that  And~XXVII is in the phase
of tidal disruption, as the RR Lyrae stars are spread all over the LBC FoV.  
However, given the proximity to M31    
we cannot rule out  that some of these RR Lyrae stars might belong to Andromeda's halo. 
 To  estimate  how many M31 halo RR Lyrae stars we can expect in the LBC FoV centered on And~XXVII 
we considered a number of different arguments. 
First of all  we assumed the surface density profile from \citet{iba2013}. Despite the complexity of the M31 halo,
\citet{iba2013} found that the azimuthally-averaged projected star-count-profile for stars
in the  metallicity range $-2.5 < [Fe/H] < -1.7$ does not have  particular
features and posses a power-law of $\Gamma =-2.3\pm 0.02$. 
 Then we considered that \citet{jeffe2011} found  five and no RR Lyrae stars in two HST/ACS field
at a distance of 35 kpc from the M31 center.
If we apply  Ibata et al. power law to the sample of RR Lyrae stars found 
 by \citet{jeffe2011} after rescaling for the different
area surveyed by our LBC images ($\sim 0.15 ~{\rm deg}^2$) and the HST/ACS  ($\sim 0.008 ~{\rm deg}^2$) 
and the different distance to the center of M31 (the projected distance to And~XXVII is 60 kpc), we find  a possible 
contamination of $\sim 14$ RR Lyrae stars from the M31 halo in 
 And~XXVII's LBC field. However, we note that 
the number of contaminant RR Lyrae stars can be larger than this estimate
as And~XXVII  lies in a region of the M31 halo where  stars in the  metallicity range $-2.5<[Fe/H]<-1.7$
form streams and are all around  And~XXVII, as  shown in Figure~9 of \citet{iba2014}. 
Thus  decontamination of And~XXVII stars  from  field objects is a difficult task.

\begin{figure}[t!]
\centering
\includegraphics[width=8.0cm,clip]{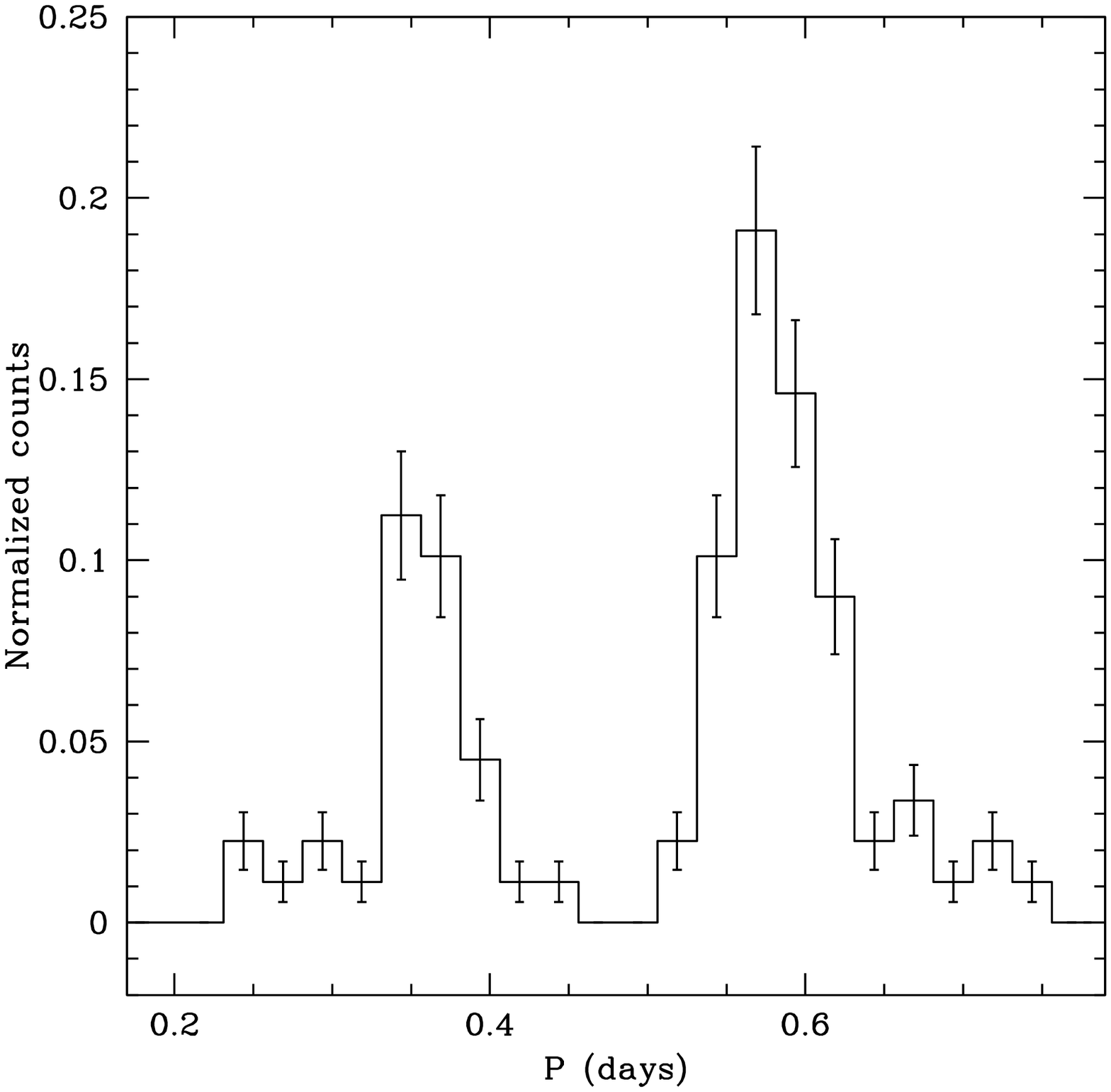}
\caption{Period distribution of  the RR Lyrae stars identified in the field of And~XXVII. 
The bin size is 0.025 days.}
\label{fig:hist}
\end{figure}

The period distribution of the RR Lyrae stars identified in the field of And~XXVII is shown by the histogram in Figure~\ref{fig:hist}.
The average period of the 58 RRab stars is  $\langle$P$_{\rm ab}\rangle$=0.59 d ($\sigma$=0.05 d),
while for the 31 RRc's is $\langle$P$_{\rm c}\rangle$=0.35 d ($\sigma$=0.04 d).
Considering that  $\sim14$  RR Lyrae stars could belong to the M31 halo we performed a bootstrap
re-sampling of the data, removing a total of 14 RR Lyrae stars. 
 Among the 14 we removed randomly 9 RRab and 5 RRc, consistently to the fraction of RRc over  total RR Lyrae stars found.
Both  average period and related $\sigma$ of RRab and RRc stars, separately, do not change significantly,
 giving us confidence that the RR Lyrae stars of And~XXVII have the 
 same properties of those  belonging to the surrounding M31 halo.

\begin{figure*}[t!]
\centering
\includegraphics[trim=25 10 0 0 clip, width=0.48\linewidth]{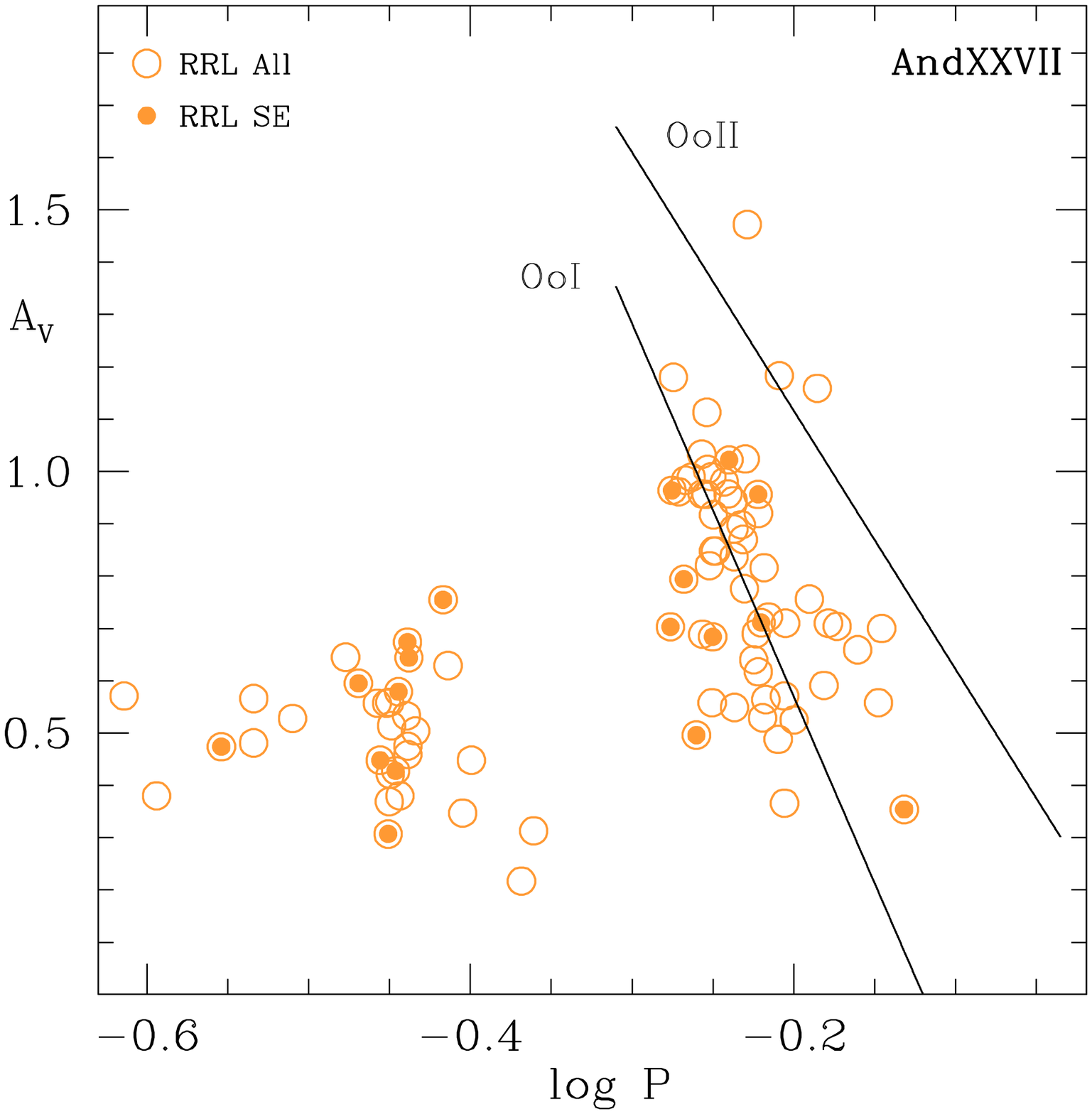}\includegraphics[trim=25 10 0 0 clip, width=0.48\linewidth]{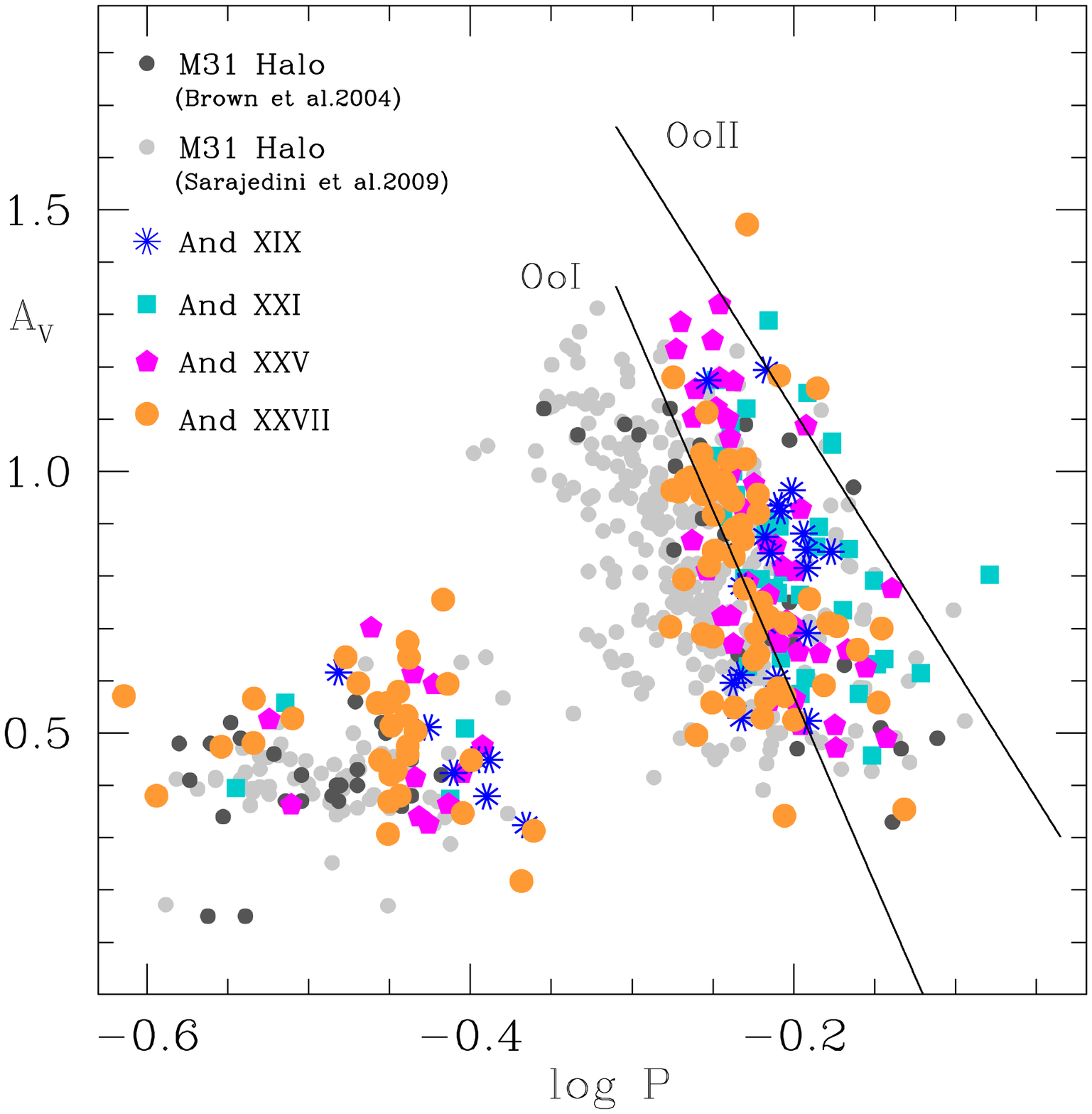}
\caption[]{ $Left$: Period-amplitude diagram of the RR Lyrae stars in the field of And~XXVII (orange open circles), 
highlighted with filled orange circles are those located in the SE region (see Section~\ref{sec:cmd}).
The two solid lines show the  loci defined by  Oo~I and Oo~II RR Lyrae stars according 
to Clement \& Rowe  (2000). 
$Right$: Period-amplitude diagram of the RR Lyrae stars in the field of And~XXVII (orange filled circles) 
 compared with the RR Lyrae stars in And~XIX (blue asterisks, from Paper~I), And~XXI (cyan squares, from Paper~II)),
 And~XXV (magenta pentagons, from Paper~III)) and in three HST fields in the M31 halo 
 from Sarajedini et al. (2009; gray dots) and Brown et al. (2004, dark-gray dots), respectively.}
\label{fig:bayl}
\end{figure*}

Based on the average period of the fundamental-mode pulsators, And~XXVII would be classified as an  Oosterhoff-Intermediate (Oo~Int)/OoI system (\citealt{oos39,cat09}),
however, we note that And~XXVII has the lowest value of $\langle$P$_{\rm ab}\rangle$   among the four  M31
satellites we have investigated  so far based on LBT data. 
The fraction of RRc to total
number of RR Lyrae stars is f$_c$= N$_c$/N$_{\rm ab+c}=0.35\pm0.07$. 
This  is much closer to the value expected for Oosterhoff~II   (Oo~II; f$_c\sim0.44$) 
than Oosterhoff~I (Oo~I; f$_c\sim0.17$) systems \citep{cat09}.
 Even in the extreme case that 5 RRc variables belong to the M31 halo and all the RRab stars to And~XXVII  
the fraction  would  become f$_c$=0.30 that  is still rather high for an Oo~I system. 
Hence,  while the average  period of the RRab stars suggests a classification as Oo-Int/Oo~I type,  the 
RRc fraction would suggest a classification more similar to Oo~II type.
The left panel of Figure~\ref{fig:bayl} shows  the period-amplitude diagram (also known as Bailey diagram, \citealt{Bai1902}) 
of the RR Lyrae stars in And~XXVII. Solid lines  are the loci defined by  RR Lyrae stars in the Oo~I Galactic globular cluster M3 (lower line) 
and the Oo~II  globular cluster $\omega$ Cen (upper line),  according to \citet{cle00}.
The majority of the RR Lyrae stars in And~XXVII are placed near the locus of the Oo I~systems and there are no differences
among RR Lyrae stars located in different parts of the LBC FoV 
(see Sections~\ref{sec:cmd}, \ref{sec:dist} and Table~\ref{t:sele}). 
We conclude that the $\langle$P$_{\rm ab}\rangle$ value and the Bailey diagram suggest a 
 classification as OoI/Int system for And~XXVII.
 
\subsection{Metallicity}\label{sec:met}
\begin{figure}[t!]
\centering
\includegraphics*[width=0.99\linewidth]{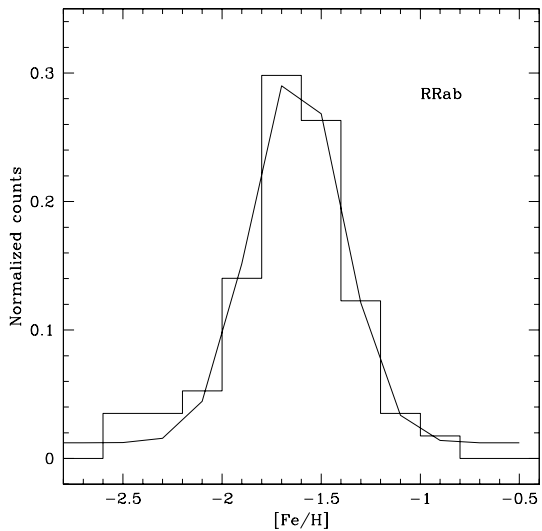}
\caption[]{Metallicity distribution of And~XXVII RRab stars obtained using Alcock et al.(2000) method. The distribution peaks at [Fe/H]$\sim -1.62$
with $\sigma \sim 0.23$ dex.}
\label{fig:met}
\end{figure}

Metallicities for the RRab stars in And~XXVII were derived using the relation of Alcock et al. (2000; see their Equation 1). 
The resulting metallicity distribution  is shown in Figure~\ref{fig:met}.  
A gaussian fit of this distribution peaks at [Fe/H]= $-$1.62  dex ($\sigma$=0.23 dex).
This value  is consistent with the photometric estimate of [Fe/H]=$-1.7\pm0.2$ dex obtained in the discovery paper \citep{rich11} 
by isochrone fitting  of the galaxy CMD,  and is within  1 $\sigma$ from the spectroscopic estimate of [Fe/H]=$-2.1\pm0.5$ dex by  \citet{col13}.
A photometric estimate of the RR Lyrae metallicity can also be derived from the
$\phi_{31}$ parameter of the Fourier decomposition of the light
curve (see, e.g., \citealt{Simon1982}, \citealt{jurk96}, \citealt{Cacciari2005}). 
We performed a sine Fourier decomposition of the $V$-band light curves of the RRab stars in our sample  
deriving the normalized Fourier parameter $\phi_{31}$ and
the deviation  parameter $D_m$, which measures the regularity of the light
curve \citep{jurk96}. According to \citet{jurk96} a reliable metallicity can be estimated  from the $\phi_{31}$ parameter  only if the
light curve satisfies the condition $D_m<3$. 

Among the RRab stars in our sample only V67 satisfies this condition.\footnote{According to \citet{Cacciari2005} variables satisfying
the relaxed condition $D_m<5$ could also be used. There are three RR Lyrae stars with $3<D_m<5$. They provide very
scattered and unreliable metallicity values hence we did not consider them.}  The light curve of V67 
is shown in Figure~\ref{fig:lca-examples}.
We applied Equation~3 in \citet{jurk96}
and derived the metallicity ${\rm [Fe/H]_{JK96}}=-1.75\pm0.49$ dex on the Jurcsik \&
Kovacs metallicity scale, which becomes ${\rm [Fe/H]_{C09}}=-1.86\pm0.50$
dex once transformed to the \citet{Carretta2009} metallicity scale using Equation~3 from \citet{Kap2011}.
All these values are consistent within their errors with the peak value of the distribution obtained applying Alcock  et al.'s method (see Figure~\ref{fig:met}).

\subsection{THE ANOMALOUS CEPHEID}\label{sec:ac}

\begin{figure}[t!]
\centering
\includegraphics[ width=1.\linewidth]{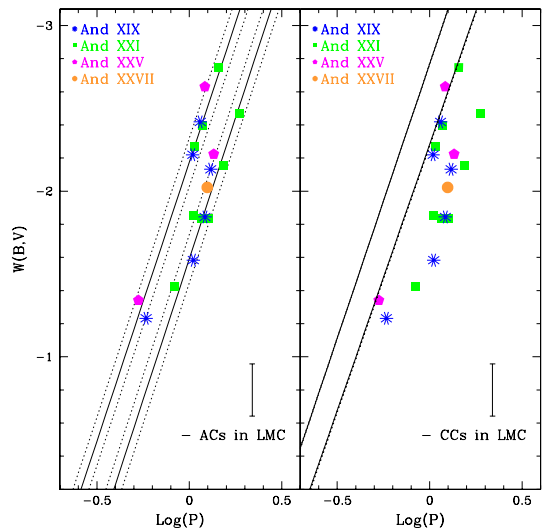}
\caption{Position on the $PW$ plane of star V2 (orange filled circle).  Solid lines represent the $PW$ relations for ACs (left panel; Ripepi et al. 2014) and CCs in 
 the LMC  (right panel;  Jacyszyn-Dobrzeniecka et al. 2016), respectively. Blue, green and magenta symbols are  ACs we have identified in And~XIX (Paper~I), And~XXI (Paper~II), and 
 And~XXV (Paper~III).
%
 The  dotted lines show the $\pm1\sigma$ deviations. For the CCs the errors of the fits are very small and the 
confidence contours are very close to the fits.}
\label{fig:pw}
\end{figure}

One of the variable stars identified in And~XXVII (V2) falls in the instability strip about one magnitude brighter than the  galaxy HB level
(see Figure~\ref{fig:cmd}). 
Following the same procedure adopted in Paper~I 
 we have compared V2  
with the period-Weseneheit ($PW$) relations of Large Magellanic 
  Cloud (LMC) ACs from \citet{ripe14}\footnote{\citet{ripe14}'s relations were derived for
    the $V$ and $I$ bands, and we have converted them to $B$ and $V$
    using equation 12 of \citet{mar04}.}  and the $PW$ relations of  LMC 
    classical Cepheids (CCs) from Jacyszyn-Dobrzeniecka et al. (2016). These comparisons are shown in the left and right panels of Figure~\ref{fig:pw}, respectively, 
    that were drawn assuming for 
    And~XXVII the distance inferred from the RR Lyrae stars (see Section~\ref{sec:dist}).  We have also plotted in the two panels of Figure~\ref{fig:pw}  
    the  ACs we have identified in And~XIX, And~XXI, and And~XXV.  

Figure~\ref{fig:pw} shows that V2 well follows the $PW$ relation for fundamental mode ACs, while it
can be definitely ruled out that the star is a short period CC (see right panel of Figure~\ref{fig:pw}). 
 The AC in And~XXVII   can be interpreted as the result of a merging in a binary system as old as the RR Lyrae stars in which mass transfer acted in the last 1-2 Gyr.
The single young star   scenario can be excluded considering that in the CMD there is no evidence of a young stellar component (see Section~\ref{sec:cmd}).

\begin{figure}[t!]
\centering
\includegraphics[trim = 20 0 0 0 clip, width=1.00\linewidth]{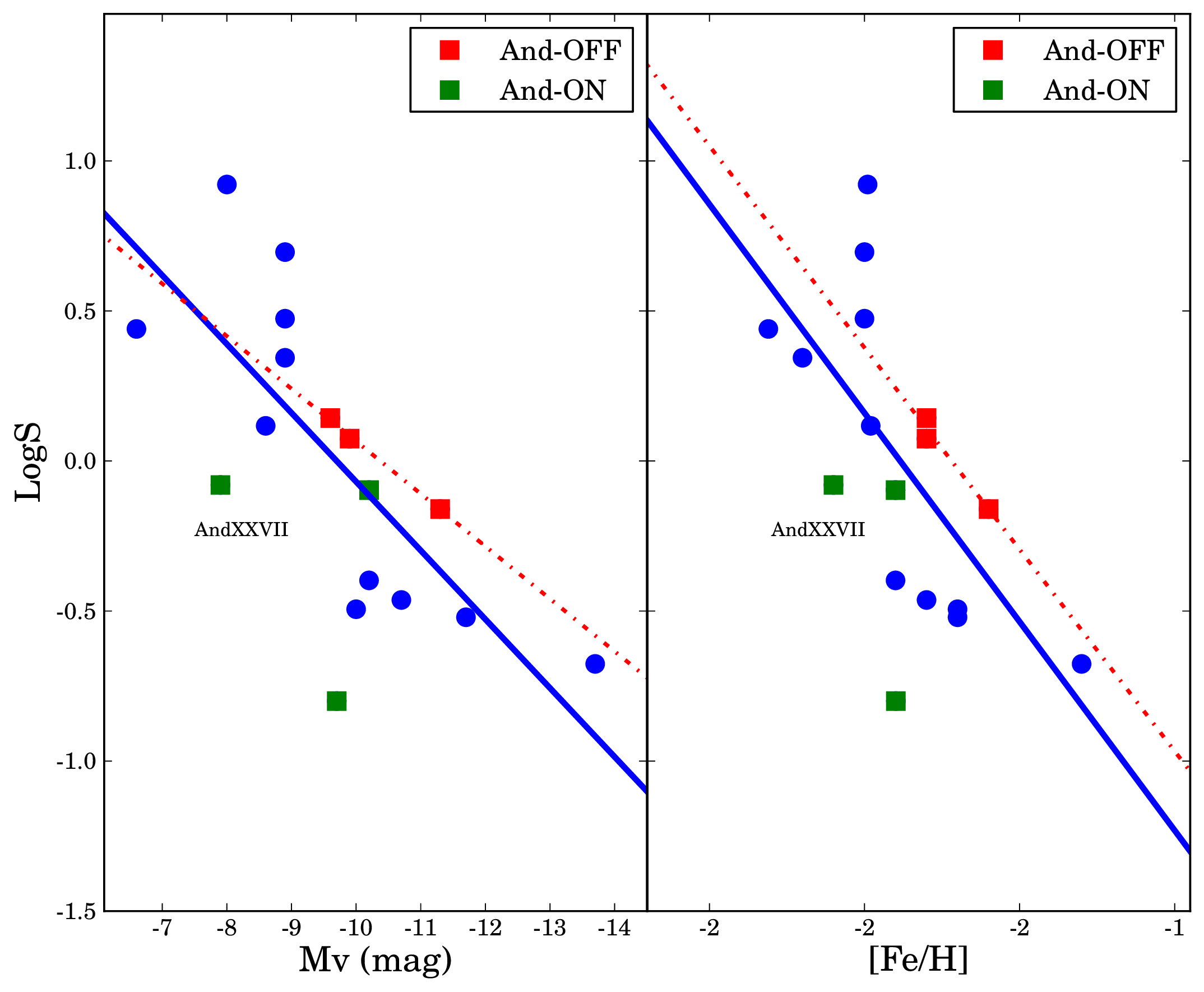}
\caption{Specific frequency of ACs in dwarf satellite galaxies of M31 and the MW vs. luminosity (left panel) and 
metallicity (right panel) of the host systems. Blue filled circles are 
MW dwarfs.  Red and green filled square are M31 satellites off and on the GPoA, respectively.}
\label{fig:specifi}
\end{figure}

As in previous papers of this series, we computed  
the specific frequency of ACs in And~XXVII assuming that  V2  is the only AC in the galaxy and compared it with  results for other galaxies. 
This is shown in Figure~\ref{fig:specifi}, where the AC specific frequencies of other Milky Way (MW; blue filled circles) and M31 (green and red filled squares) satellites 
are taken from Paper~III. The blue ridge line shows the relation obtained from all the dwarf satellites in the plot. The M31 satellites that are off the GPoA (red squares) seem 
to follow a different trend shown by the red dashed line. The sample is still statistically too poor to conclude that off- and on-plane satellites 
obey to different correlations between ACs specific frequency and luminosity or metallicity, but the satellites studied so far seem to suggest that this may be the case. 

\section{CMD AND PROJECTED SPATIAL DISTRIBUTIONS }\label{sec:cmd}

\begin{figure*}[t!]
\centering
\includegraphics[width=0.6\linewidth]{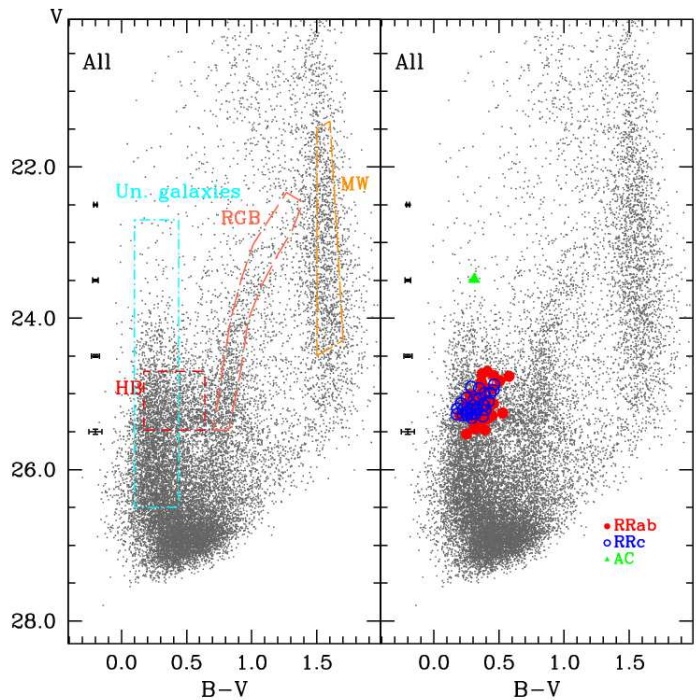}
\caption[]{$Left$: CMD of the sources in the whole LBC FoV. Only objects with $-0.35 \leq$ Sharpness $\leq$0.35 and $\chi < $1.5 are displayed. 
 The RGB, HB and MW selections are marked with dashed regions. $Right$:
Same as left but with superimposed the variable stars. Red filled circles are the RRab stars, blue open circles are the RRc stars 
and the green triangle is the  AC.}
\label{fig:cmd}
\end{figure*}

\begin{figure}[t!]
\centering
\includegraphics[trim = 75 0 40 0 clip, width=1.0 \linewidth]{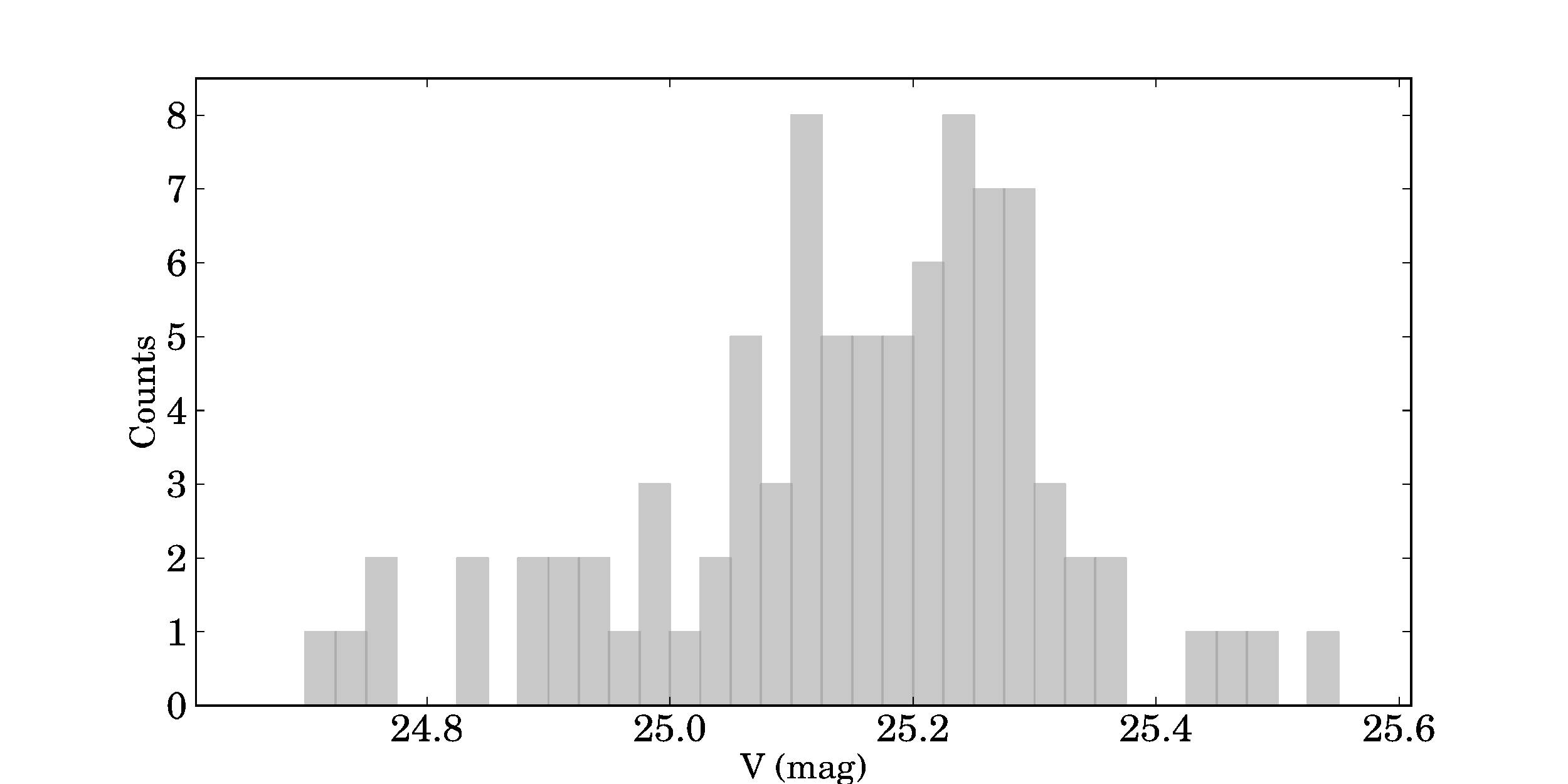}
\caption{Distribution in apparent  $V$ mean magnitude of the 89 RR Lyrae stars identified in the field of And~XXVII. 
The bin size is 0.025 mag.}
\label{fig:histo-all}
\end{figure}

The CMD of the sources in the whole LBC FoV 
 is presented in Figure~\ref{fig:cmd}. Only sources with DAOPHOT quality image parameters: $-0.35 \leq$ Sharpness $\leq$0.35 and $\chi < $1.5 are displayed. 
The CMD appears to be heavily contaminated by MW foreground stars in the red part ($1.4\le B-V \le 1.9$ mag)  and by  background 
unresolved galaxies in the blue part   \citep[$-0.1\le B-V \le 0.1$ mag, see Section 9 in Paper~I and e.g.][]{belli2010, bella2011}. 
In the  right panel plot, red filled circles are RRab stars  (58 sources), blue open circles are  RRc stars (31 sources) 
and   the green triangle is the   AC.
The RR Lyrae stars trace the HB,  their distribution in apparent $V$ mean magnitude  
is very broad ranging from  24.70 to 25.55 mag (see Figure~\ref{fig:histo-all}). A most significant peak  
is observed  at $\langle V(RR)\rangle \sim$ 25.25 mag, followed by a second peak at $\langle V(RR)\rangle \sim$ 25.10 mag.

\begin{figure*}[t!]
\centering
\includegraphics[width=1\linewidth]{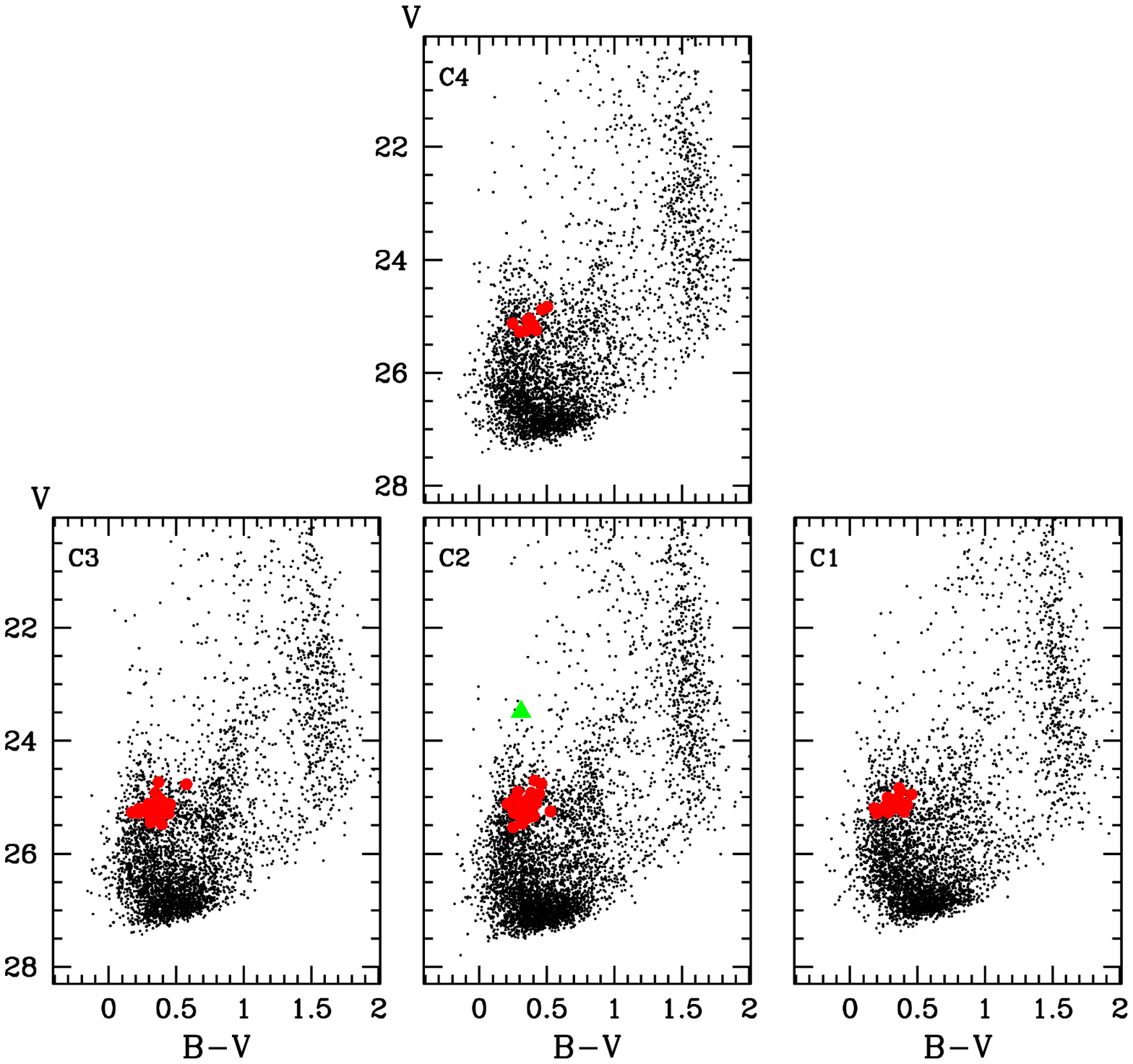}
\caption[]{CMDs of sources in the 4 CCDs of the LBC mosaic. Only objects with $-0.35 \leq$ Sharpness $\leq$0.35 and $\chi < $1.5 are displayed. 
 Red filled circles are the RR Lyrae stars: 32 in C3, 29 in C2, 21 in C1 and 7 in C4. The green triangle is the  AC which is located in C2.}
\label{fig:cmd-4pan}
\end{figure*}

\begin{figure*}[t!]
\centering
\includegraphics[trim = 85 0 0 0 clip, width=1.10 \linewidth]{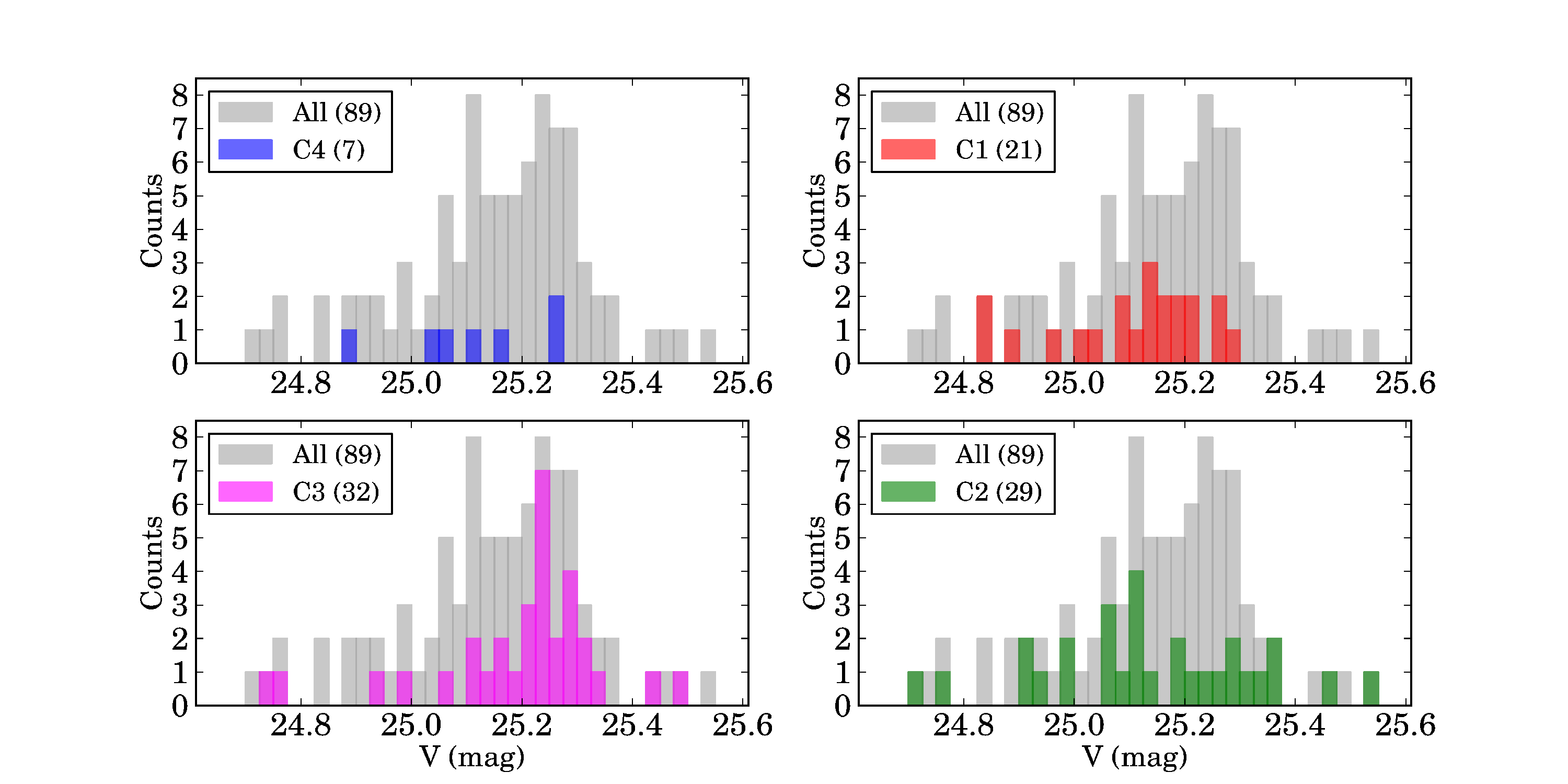}
\caption{Distribution in apparent  $V$ mean magnitude of the RR Lyrae stars in the 4 CCDs of the LBC mosaic. 
The bin size is 0.025 mag.}
\label{fig:histo-4ccd}
\end{figure*}

\begin{figure*}[t!]
\centering
\includegraphics*[trim=25 10 0 0 clip, width=0.48\linewidth]{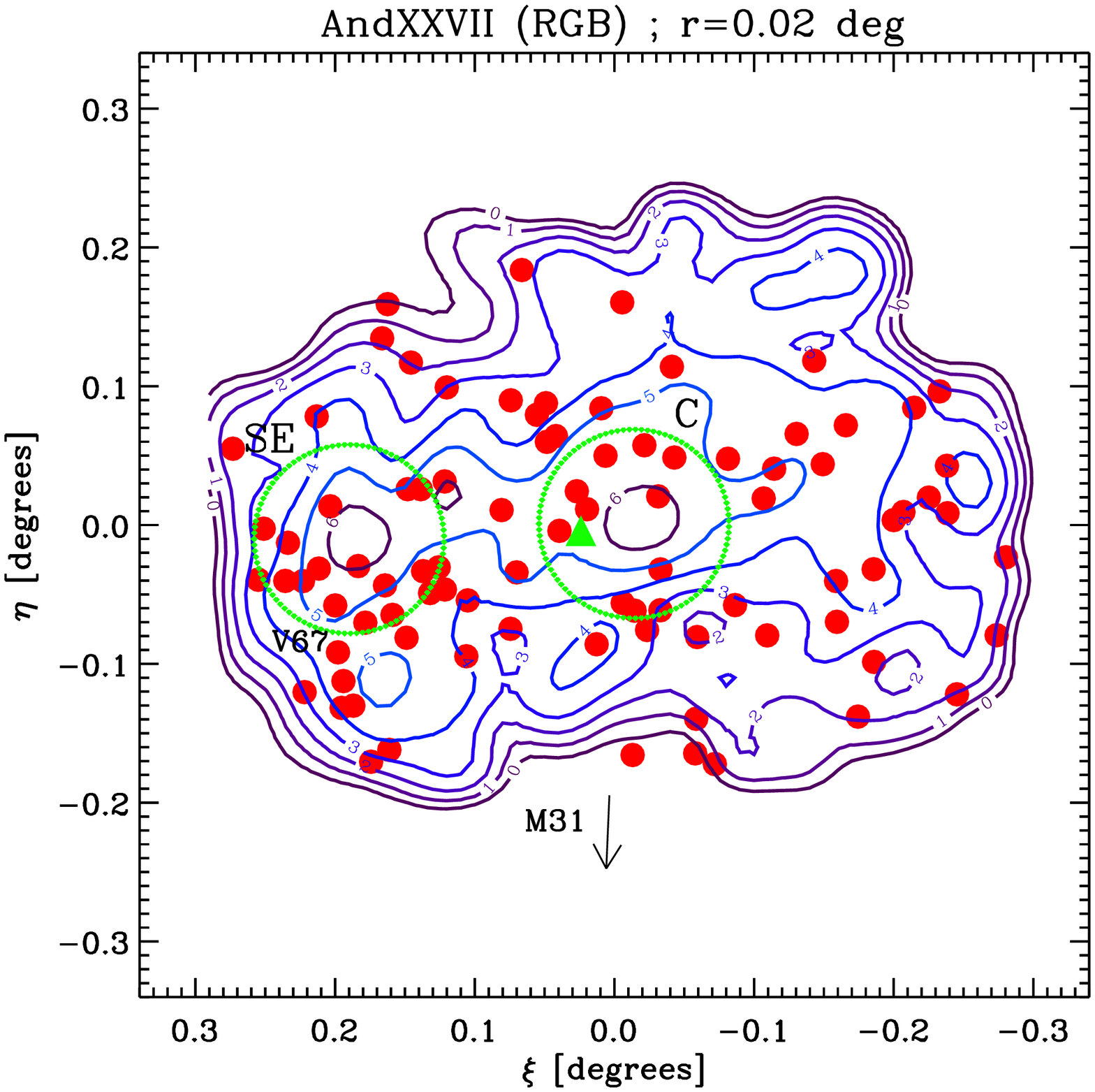}\includegraphics*[trim=25 10 0 0 clip, width=0.48\linewidth]{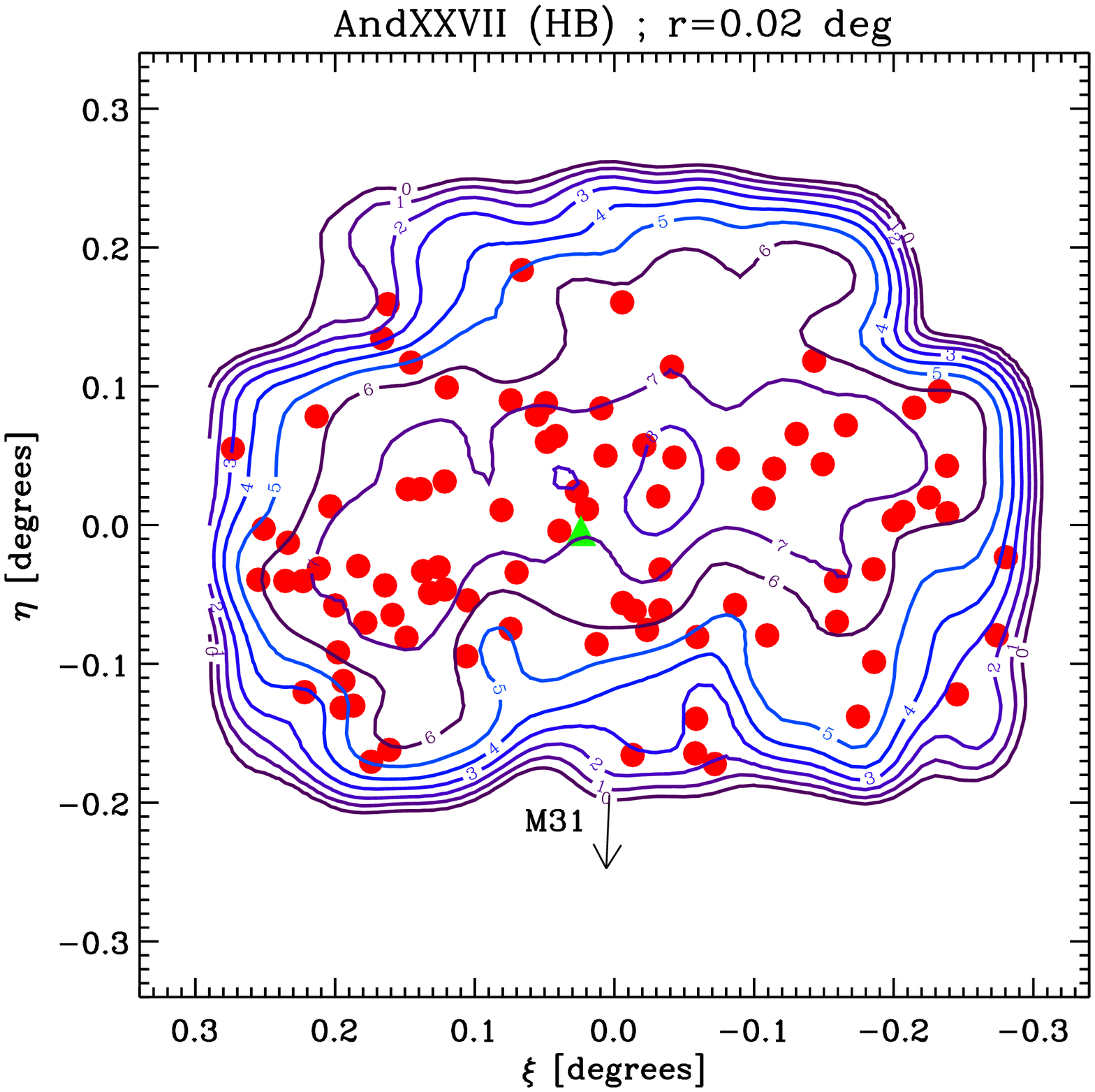}
\caption[]{$Left$: Isodensity contours of RGB stars in And~XXVII. X and Y axes are, respectively, RA and decl.  
differential coordinates computed from the center of the galaxy given by Richardson et al. (2011).
Red filled circles are  RR Lyrae stars, the  green filled triangle is the AC. We have labeled the RR Lyrae star V67 for which we have estimated a metallicity 
of [Fe/H]=$-1.86 \sim$ 0.50 dex (on the \citealt{Carretta2009} scale)  from the Fourier parameter $\phi_{31}$. 
The green circles mark  areas of 4 arcmin in radius around regions C and SE, respectively.
$Right$: Same as in the left panel but for And~XXVII HB stars. 
}
\label{fig:isoden}
\end{figure*}

The four panels in Figure~\ref{fig:cmd-4pan} show the CMDs of sources in the 4 individual CCDs of the LBC mosaic (labeled C1, C2, C3 and C4).  
Similarly Figure~\ref{fig:histo-4ccd} shows the distributions in apparent $V$ mean magnitude  of the RR Lyrae stars  in C1, C2, C3 and C4, separately.
Although contamination by foreground MW stars/background unresolved galaxies (see left panel of Figure~\ref{fig:cmd}) and broadening of both RGB and HB is 
present in each of the four individual CMDs, C3 shows a more prominent RGB extending as bright as $V \sim$ 22.5 mag, whereas the RGB in the other 3 CCDs 
seems to be truncated around $V \sim$ 23.5 mag. Furthermore, as shown by Figure~\ref{fig:histo-4ccd}
the peak at $\langle V(RR)\rangle \sim$ 25.25 mag is mainly due to the RR Lyrae stars located in C3.\\
To  investigate the spatial distribution of And~XXVII stars we selected objects in the 
RGB region of the CMD obtained in the total LBC FoV  (see left panel of Figure~\ref{fig:cmd}) with magnitude and color in the ranges of $ 22.5\le V \le25.3$ mag
and  $  0.8\le B-V \le1.3$ mag,  respectively.
We built isodensity maps by  binning these RGB  stars in $1.2'\times1.2'$ boxes and smoothing with a Gaussian kernel of
$1.2'(0.02^{\rm{o}})$ FWHM. 
The left panel of Figure~\ref{fig:isoden} shows the RGB isodensity maps 
where the outermost  contour levels are $3\sigma$ above the sky background.
Interesting features are revealed by the isodensities contours  of this selection.
 A high-counts isodensity  contour shows up around \citet{rich11}'s  center coordinates of And~XXVII. This region, named C in the figure,
 has been marked with a  dotted circle of 4 arcmin in radius that is entirely contained into C2.  
The second isodensity  has a very elongated structure that extends in south-east direction and reaches a second high-counts isodensity displaced by about 0.2 degrees
in south-east direction from the center of the C region. This second region has been named SE and  is marked by a dotted circle of 4 arcmin in radius
with center coordinates: R.A.=$00^{\rm h}38^{\rm m}10.4^{\rm s}$, decl.=+45$^{\circ}21{\arcmin}34\arcsec$ that is entirely contained into C3.  
In both regions, SE and C, we adopted  a radius of 4 arcmin that defines an area twice the high-counts isodensity contours.
In the SE region there is a concentration of stars as high as  in the C region and a higher number of RR 
Lyrae stars. 

\begin{figure}[t!]
\centering
\includegraphics[width=1.\linewidth]{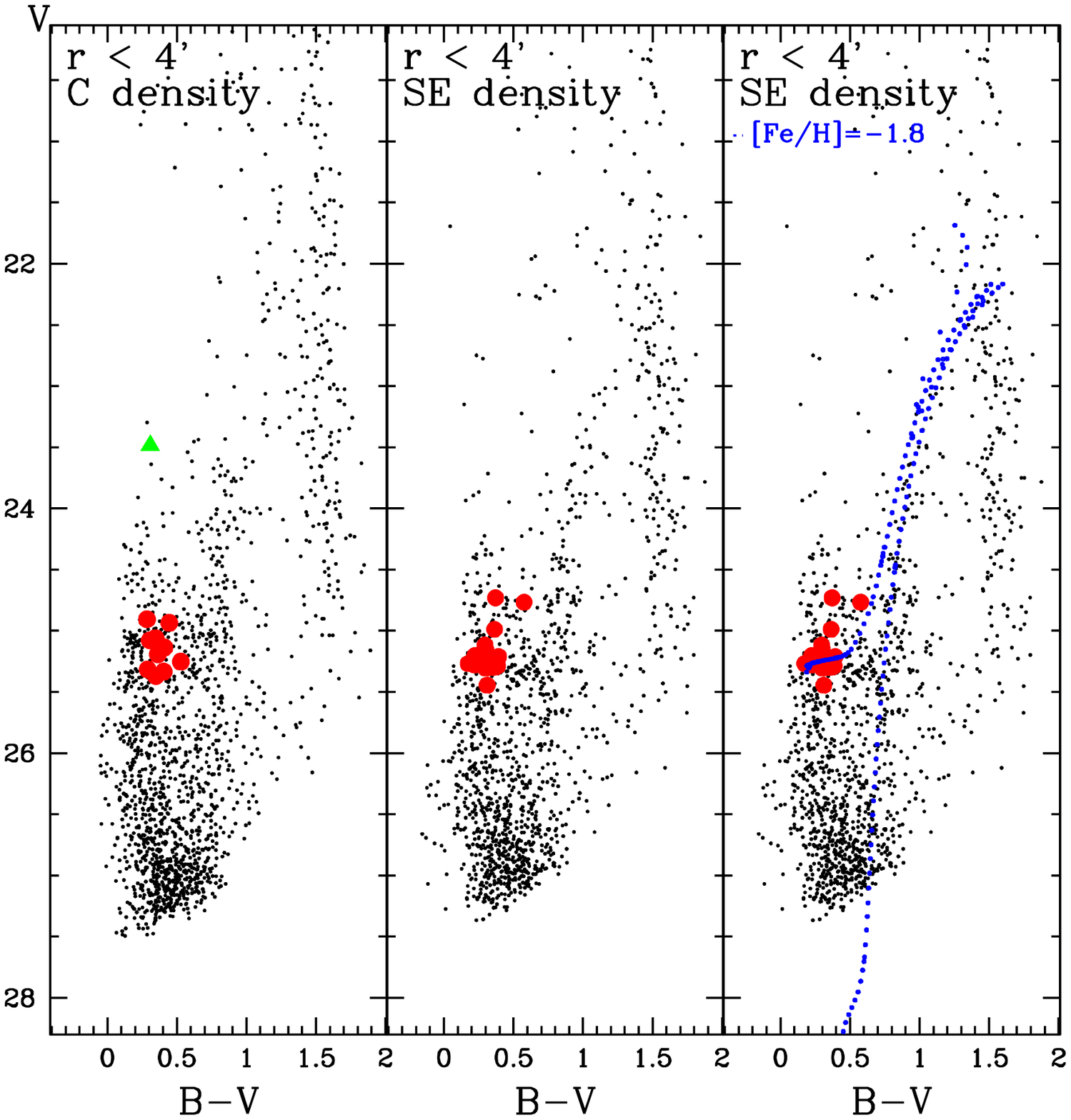}
\caption[]{$Left$: CMD of sources in the C  overdensity region.  $Center$: Same as in the left panel, but for the SE region. $Right$: 
CMD of the SE region with superimposed  an isochrone of 13 Gyr with [Fe/H]=$-1.8$ dex (blue dotted line; see text for details).}
\label{fig:bicmd}
\end{figure}

\begin{figure}[t!]
\centering
\includegraphics[trim=50 0 30 0, width=1.0\linewidth]{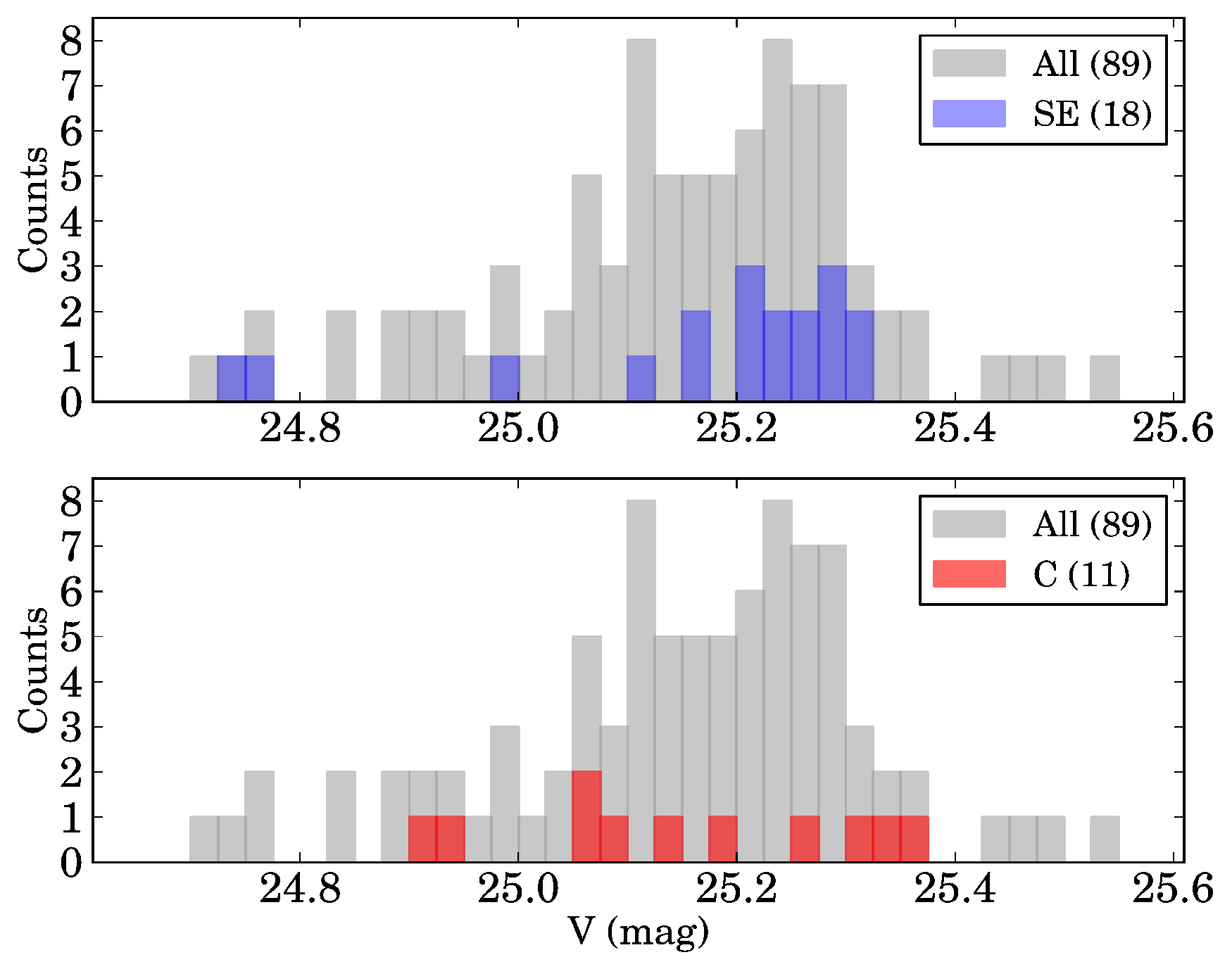}
\caption[]{$Lower~panel$: Distribution in apparent $V$ mean magnitude of the 11 RR Lyrae stars in the  C region. $Upper~panel$: Same as in the lower panel, but for the 18 RR Lyrae stars in the SE region. }
\label{fig:bihisto}
\end{figure}

 The presence of two high density contours in the map of Figure~\ref{fig:isoden} points out that
And~XXVII has a very complex physical structure.
 We have to stress that the contamination of background and foreground sources can hamper the 
RGB stellar counts even if the selection was performed in the specific region of the CMD containing RGB stars 
at the distance of And~XXVII (see left panel of Figure~\ref{fig:cmd}).
Moreover, a  complex physical structure of And~XXVII  is also corroborated by the wide RGB and the large spread  in mean apparent  magnitude 
of the RR Lyrae stars ($\langle V(RR)\rangle$) that can be 
attributed a distance spread.
 Three possible alternative scenarios to explain  And~XXVII's structure and its two stellar concentrations are presented  in Section~\ref{concl}. 
  Isodensity contours were also computed selecting stars in the HB region ($ 25.6\le V \le24.6$ mag, $  0.17\le B-V \le0.64$ mag) of And~XXVII CMD, 
  as shown in the right panel of Figure~\ref{fig:isoden}.
  The HB region of the CMD is contaminated by unresolved background galaxies  (blue box,  left panel of Figure~\ref{fig:cmd}), 
  but still  the 
   HB stars well follow the structure traced by  the RGB stars.
    We tried to select And~XXVII's HB stars choosing  objects closest to the CMD region defined by the RR Lyrae stars, 
    but given the presence in this region of the CMD of unresolved background galaxies, the exact shape of the HB isodensities depends on the selection performed. 
    For the RGB selection the contamination
   is not as crucial as for HB stars, indeed
   if we make multiple selections of the RGB region the shape of the isodensity contours does not 
   change significantly. 
 In order to characterize the stellar populations in the 
   two isodensities with the highest stellar counts we drew the CMDs of the sources in the C and SE regions separately. 
 The result is shown in Figure~\ref{fig:bicmd}. The CMD of the SE region is similar to the CMD of C3 and shows a well extended, 
 better defined and narrower RGB especially around the tip,  than the CMD of the C region which RGB appears instead truncated around $V \sim 23.5$ mag 
 as also is the CMD of C2. The HB of SE is also better defined than that in region C. The lower panel of Figure~\ref{fig:bihisto} 
 shows the distribution in apparent $V$ mean magnitude of the 11 RR Lyrae stars 
in region C, while the upper panel shows the mean magnitude distribution  of the 18 RR Lyrae stars in region SE. 
Fifteen of the 18 RR Lyrae stars contained in SE trace a rather tight HB with 
$\langle V(RR)\rangle$ = 25.24 mag and  $\sigma$= 0.06 mag (average on 15 stars).  On the contrary, the 11 RR Lyrae stars of region C  are rather spread
in magnitude (see left panel of Figure~\ref{fig:bicmd} and bottom panel of Figure~\ref{fig:bihisto}) with $\langle V(RR)\rangle$ = 25.15 mag and  
$\sigma$= 0.15 mag. This makes us wonder whether the actual center of And~XXVII might be in the SE region rather than in C, as found by \cite{rich11}.

We have superimposed on the SE CMD  an isochrone  of 13 Gyr with [Fe/H]=-1.8 as derived from the web  interface CMD 2.9\footnote{$http://stev.oapd.inaf.it/cgi-bin/cmd$}
based on the \citet{marigo2017} evolutionary tracks. The isochrone was corrected for the  distance and   reddening derived in Section \ref{sec:dist} for the SE region.
This isochrone well fits the position of  RR Lyrae and RGB stars. 
The comparison between  isochrone and observed CMD suggests  a higher metallicity value
for And~XXVII  than  measured by \citet{col13}. The isochrone fit result is consistent (within the errors) 
 with the metallicity estimated using the RR Lyrae stars in Section~\ref{sec:met}.  
A similar conclusion 
is also reached by comparison
 with the RGB and HB ridge lines of galactic Globular Clusters (GCs) with different metallicities,  the 
best result is obtained for NGC 4147 with [Fe/H]=-1.8 dex \citep{harr96}.
In the following we adopt [Fe/H]=$-1.8$ dex for the metallicity of And~XXVII. 

\section{DISTANCE AND  STRUCTURE}\label{sec:dist}
The mean luminosity of the RR Lyrae stars can be used to estimate the distance 
to And~XXVII.  We are aware that some of the RR Lyrae stars in the field of And~XXVII
can belong to the M31 halo, but since it is not possible to distinguish them, 
in deriving the distance we used all the sample of 89 RR Lyrae stars found in this work. 
The average $V$ apparent  magnitude of  the 89 RR Lyrae stars  in the field of And~XXVII is $\langle V(RR) \rangle$=25.15 mag ($\sigma$=0.17 mag average on 89 stars). 
As in previous papers of this series we derived the reddening from the  RR Lyrae stars  
adopting the method of \citet[][equation at page 1538]{pier02}. The reddening estimated in this way,  $\langle E(B-V) \rangle=0.04\pm0.05$ mag 
 (where the error is the standard deviation of the mean), is slightly smaller  but still consistent within 1 $\sigma$ 
 with the value inferred by \citet[][$E(B-V)=0.08\pm0.06$ mag]{sch98}. The visual absorption A$_V$ was derived using 
 the extinction law A$_V=3.1\times E(B-V)$ from \citet{card1989}.
We then adopt  M$_{\rm V}=0.54\pm0.09$ mag  
for the absolute visual magnitude of RR Lyrae stars with metallicity  [Fe/H] = $-$1.5 dex \citep{cle03} and
$\frac{\Delta {\rm M_V}}{\Delta {\rm[Fe/H]}}=-0.214\pm0.047$ mag/dex \citep{cle03,gra04} for the 
slope of the RR Lyrae luminosity-metallicity relation. For the metallicity of And~XXVII we adopt [Fe/H]=$-1.8\pm0.3$ dex
as derived from isochrone fitting and consistent with the estimate from the RR Lyrae stars.  
The resulting distance modulus is (m-M)$_0$=$24.55\pm0.26$ mag. This is 0.16 mag fainter but still consistent, within the large error,  
with \cite{rich11}  lower limit of $\geq$ 757 $\pm$ 45  kpc (corresponding to (m-M)$_0$=$24.39\pm0.13$ mag) and placing And~XXVII within  the M31 complex.
The  faintest estimate by \citet{con12}: (m-M)$_0$=$25.49^{+0.07}_{-1.03}$  mag has a very large asymmetric error likely due to the difficulty in identifying 
 the tip of RGB which is very scarcely populated in And~XXVII. In any case, our estimate is consistent within the errors with the lower limit of  \citet{con12} estimate.
 The large error in our distance estimate is due in a good fraction to the significant dispersion of the average magnitude of the whole RR Lyrae star sample. 
 As anticipated  in Section~\ref{sec:cmd},  And~XXVII  appears to be in the process of tidal disruption, 
 this may have resulted in the RR Lyrae stars to be located at different distances from us, thus 
causing the large dispersion in  $\langle V(RR) \rangle$.
 To further investigate  the galaxy structure 
we have compared number and properties of the RR Lyrae stars located in different parts of the LBC FoV starting from the two regions with highest stellar counts, namely,
regions C and SE in Figure~\ref{fig:isoden}.  


\begin{figure}[t!]
\centering
\includegraphics[trim=25 0 25 0, width=1\linewidth]{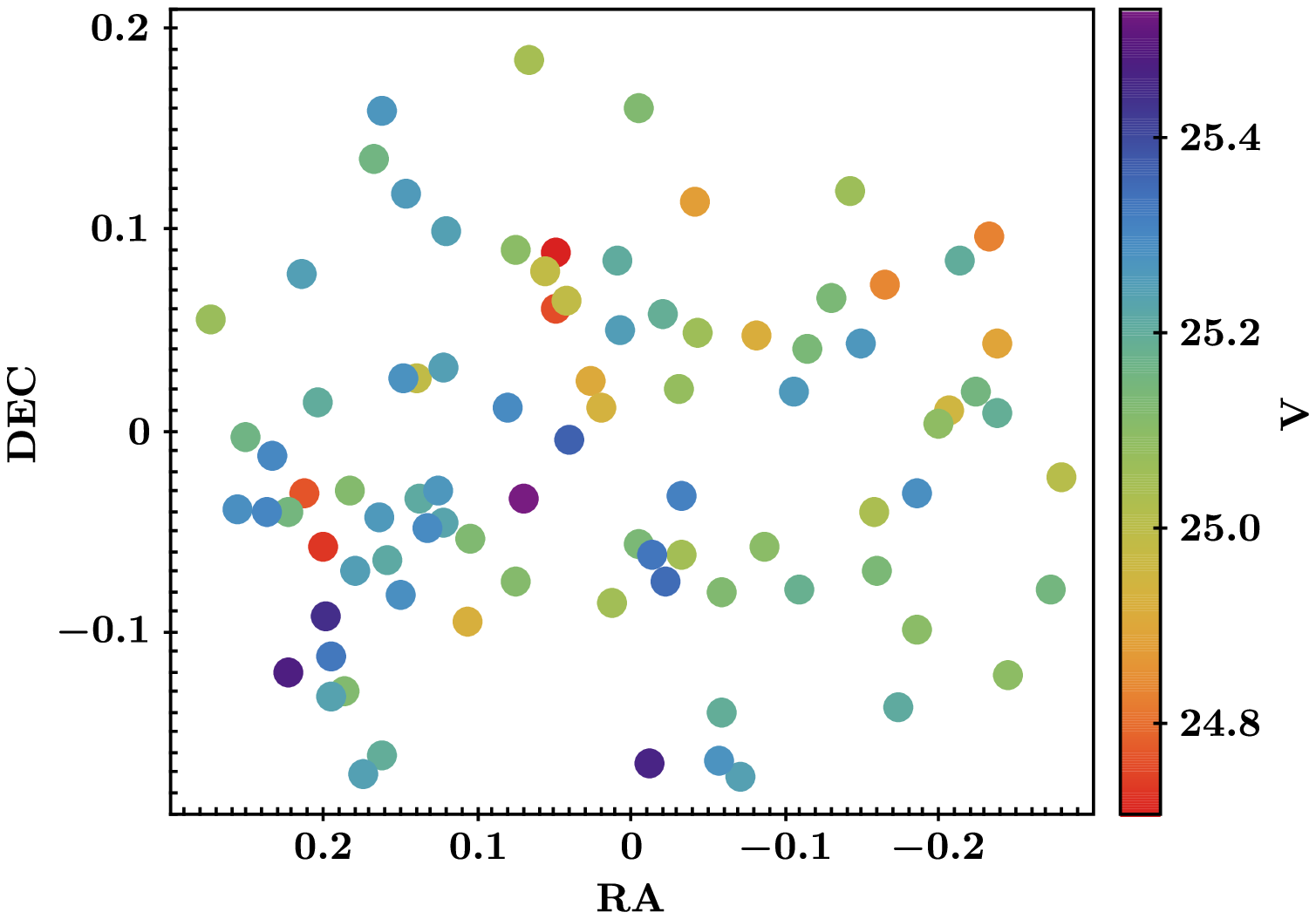}
\caption[]{Spatial distribution of the RR Lyrae stars in the field of And~XXVII. Stars are color-coded according to their  mean $V$ magnitude.}
\label{fig:colormap}
\end{figure}
There are 18 RR Lyrae stars in the SE region and only 11 in the C region. This difference remains even if we reduce the radius of the two regions: 
in a circular portion of SE  of 3 arcmin in radius there are 8 RR Lyrae stars to compare with  6 RR Lyrae  contained in a same portion of  region C.
Average $V$ magnitude,  reddening and other characteristics (total number of RR Lyrae stars, average RRab period, distance modulus and fraction of RRc pulsators)
for the RR Lyrae stars in region C  and SE separately  
are provided in Table~\ref{t:sele}. Two sets of values are listed for region SE, the first one corresponding to the whole set  of 18 variables, 
and the second one (labeled SE1 in the table) considering only the 15 RR Lyrae stars whose mean magnitude peaks around  $\langle V \rangle \sim$ 25.24 mag. 
Distance moduli of C and SE differ by 0.08 mag in the first case, this difference is still within 1 $\sigma$ given the large dispersion of the estimates.
The difference is of 0.21 mag in the second case (SE1) and may be due to a  real radial distance effect.
 However,  considering the large errors  we are dealing with,   this conclusion has to be taken with caution. The spread
in magnitude can be due to a real three-dimensional (3D) effect, but not as big as 0.21 mag.
The centers of regions C and SE are $\sim$ 0.2 degree a part in the sky, that at the distance 
of And~XXVII corresponds to a physical projected separation of $\sim$ 3 kpc.
Is this an indication that
the galaxy is disrupted and the SE component is 3 kpc away from the galaxy center? Or, since SE has the cleanest CMD and the largest concentration of RR Lyrae stars  
most of which are located approximately at the same distance from us, could SE be the remaining nucleus of the now disrupted And~XXVII?

\begin{table*}
\caption[]{Properties of selected samples of RR Lyrae stars  in And~XXVII. }
\scriptsize
\label{t:sele}
\begin{center}
\begin{tabular}{l c c l l l c c  }
\hline
\hline
\noalign{\smallskip}
id       &   N (RRab+RRc) &     $\langle V(RR) \rangle$    & $\langle$P$_{\rm ab}\rangle$         & $\langle$P$_{\rm c}\rangle$       &  E(B-V)      & (m-M)$_0$      & f$_c$     \\
\noalign{\smallskip}
\hline
\noalign{\smallskip}
ALL      &   58+31        & $25.15\pm0.17$ & $0.59\pm0.05$ & $0.35\pm0.04$ & $0.04\pm0.05$ & $24.55\pm0.26$   & 0.35   \\
SE &   9+9                & $25.17\pm0.17$ & $0.58\pm0.05$ & $0.35\pm0.03$ & $0.05\pm0.05$ & $24.54\pm0.26$   & 0.50   \\
SE1 &   6+9              & $25.24\pm0.06$ & $0.55\pm0.01$ & $0.35\pm0.03$ & $0.03\pm0.05$ & $24.67\pm0.20$   & 0.60  \\
C   &    6+5              & $25.15\pm0.15$ & $0.62\pm0.03$ & $0.33\pm0.05$ & $0.07\pm0.05$ & $24.46\pm0.25$   & 0.45   \\
NORTH    &    29+15       & $25.09\pm0.16$ & $0.60\pm0.04$ & $0.36\pm0.04$ & $0.05\pm0.03$ & $24.46\pm0.22$   & 0.35    \\
SOUTH    &    29+16       & $25.20\pm0.15$ & $0.59\pm0.05$ & $0.34\pm0.04$ & $0.03\pm0.06$ & $24.63\pm0.27$   & 0.35   \\
 \hline 
\end{tabular}
\end{center}
\normalsize
\end{table*}

Triggered by these two different possibilities we divided the RR Lyrae stars in two samples, corresponding, respectively, to the south and north portions 
of And~XXVII according to \cite{rich11} center coordinates.  
Characteristics  of the RR Lyrae stars in the north and south samples are also summarized in Table~\ref{t:sele}. 
North and south distance moduli differ by $\sim0.2$ mag, that although still consistent within the respective errors could indicate a 3D  effect where the north part of 
And~XXVII  is  closer to us than the south part.

We also tested whether possible zero point differences among the four CCDs of the LBC might cause the observed differences, 
by comparing the median apparent $B$ and $V$ magnitudes of objects in the blue part ($-0.1\le B-V \le 0.1$ mag, $24 \le V \le 26$ mag)
and in the red part ($1.6\le B-V \le 1.9$ mag, $23 \le V \le 25$ mag)
of the CMD.

Differences among the 4 CCDs range from 0.001 to 0.03 mag both in the $B$ and $V$ bands, hence cannot explain the 
difference observed between average apparent magnitudes of the north and south RR Lyrae subsamples. Similarly, these differences cannot be explained 
by differential reddening, as reddening is  generally small and very similar among the various RR Lyrae subsamples.
 We computed also the average reddening derived from the RR Lyrae stars in the four CCDs obtaining:
 $\langle E(B-V) \rangle_{C1}=0.03\pm0.06$ mag,   $\langle E(B-V) \rangle_{C2}=0.03\pm0.06$ mag,  $\langle E(B-V) \rangle_{C3}=0.05\pm0.05$ mag
and  $\langle E(B-V) \rangle_{C4}=0.01\pm0.06$ mag.
Therefore we conclude that the magnitude differences are real 
and likely due to a 3D effect. 
Figure~\ref{fig:colormap} shows the spatial distribution of the 89 RR Lyrae stars in the field of And~XXVII,
where each source is color-coded according to its mean  apparent magnitude 
($\langle V \rangle$). 
If we assume that differences in  $\langle V \rangle$ values are totally  due to a projection/distance effect, which is sensible given the similarity of pulsation 
properties of the 89 RR Lyrae stars and the low and homogeneous reddening in the area,  the distribution in Figure~\ref{fig:colormap}  suggests that  And~XXVII center 
likely is in the SE overdensity and the galaxy is tilted with the north-west portion closer 
to us than the south-east part. 

We recall that among the three other M31 satellites we observed with the LBT,  we did not find evidence of a particular smearing in mean magnitude of the RR Lyrae 
stars in And~XIX and And~XXV,
whereas the magnitude spread observed in And~XXI  was found to be fully justified by the presence of two RR Lyrae populations with different metallicities as also suggested
by the bi-modal period distribution of the RRab stars 
(see Figure~2 in Paper~II). However,  the scatter of the RR Lyrae star mean magnitude in the field of And~XXVII is over twice than observed in And~XXI, 
while there is no evidence of
bi-modality  in  period of the RRab stars. This along with the large spread in the spatial distribution of the RR Lyrae population make us to conclude 
 that either we are  resolving  the 3D structure of a galaxy in the final stage of disruption  or  we are sampling the RR Lyrae stars
in And~XXVII together with those in a background/foreground structure like  the M31 NW stream or the M31 halo.

\subsection{Different old stellar populations?}

If the differences in magnitudes of  the RR Lyrae star subsamples are not due to distance effects a possible alternative explanation is the 
presence in And~XXVII of  different old stellar generations.
To test this hypothesis the positions on the CMD of the different samples of RR Lyrae stars listed in Table~\ref{t:sele}  were 
compared to the model of \citet{marc2015} and the 13 Gyr isochrones  of \citet{marigo2017}  after correction
for the reddening  and the distance modulus derived using our whole RR Lyrae stars sample
[(m-M)$_0$=$24.55\pm0.26$ mag, $\langle E(B-V) \rangle=0.04\pm0.05$ mag].
To further constrain the observations we compared  also the observed period distribution of the RRab pulsators to the models.
Starting from the C sample the position on the CMD and the period distribution of the RRab can be reproduced by the models
with metallicity from $[Fe/H]=-2.3$ to $-1.5$ dex, M=0.67-0.8 M$_\odot$, T${_eff}$=5900-6900 K and LogL=1.69-1.79 L$_\odot$.
The  $[Fe/H]=-1.5$  models are at the limit of the redder RR Lyrae stars. 

\begin{figure*}[t!]
\centering
\includegraphics*[trim=25 10 0 0 clip, width=0.48\linewidth]{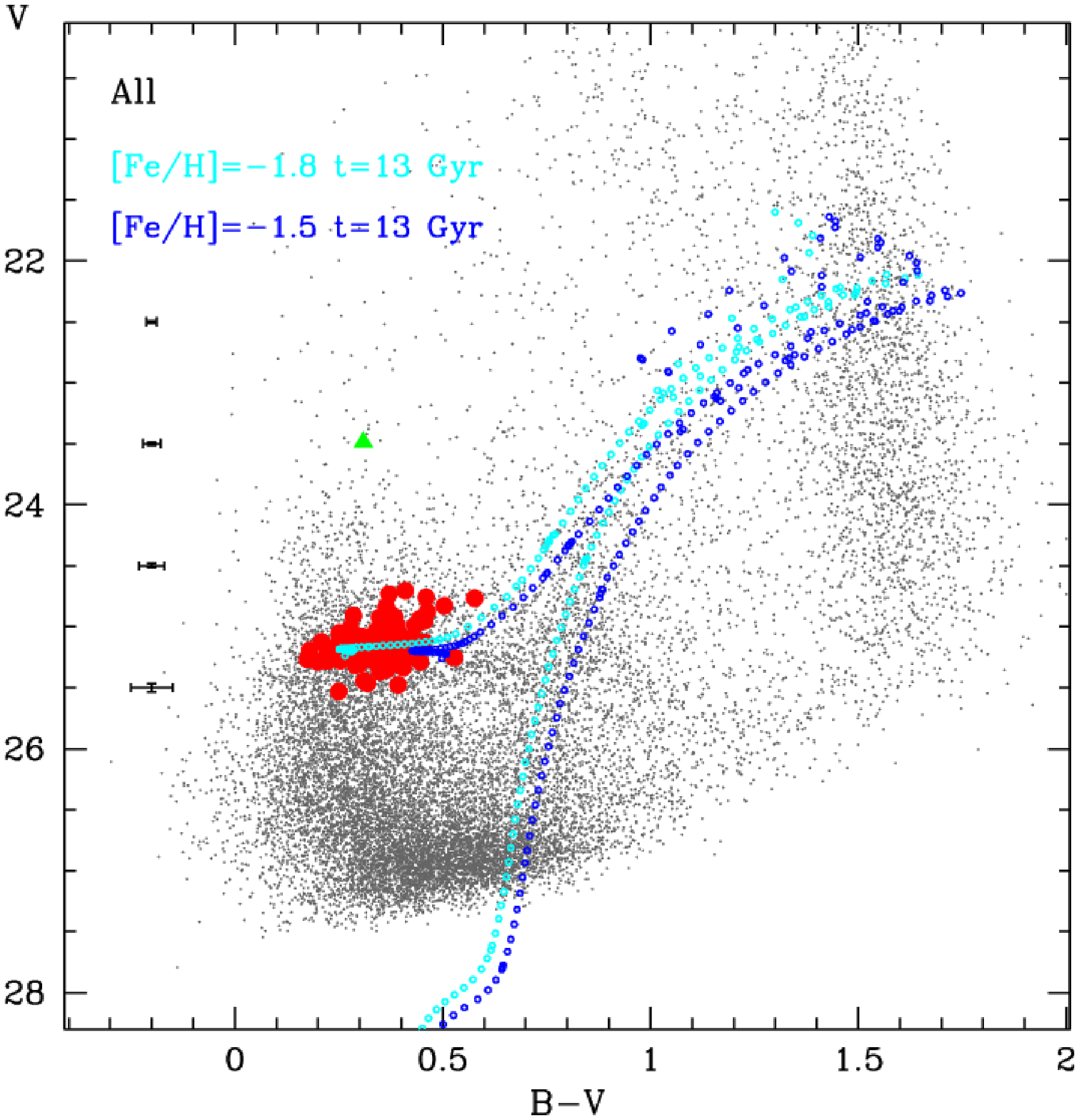}\includegraphics*[trim=25 10 0 0 clip, width=0.48\linewidth]{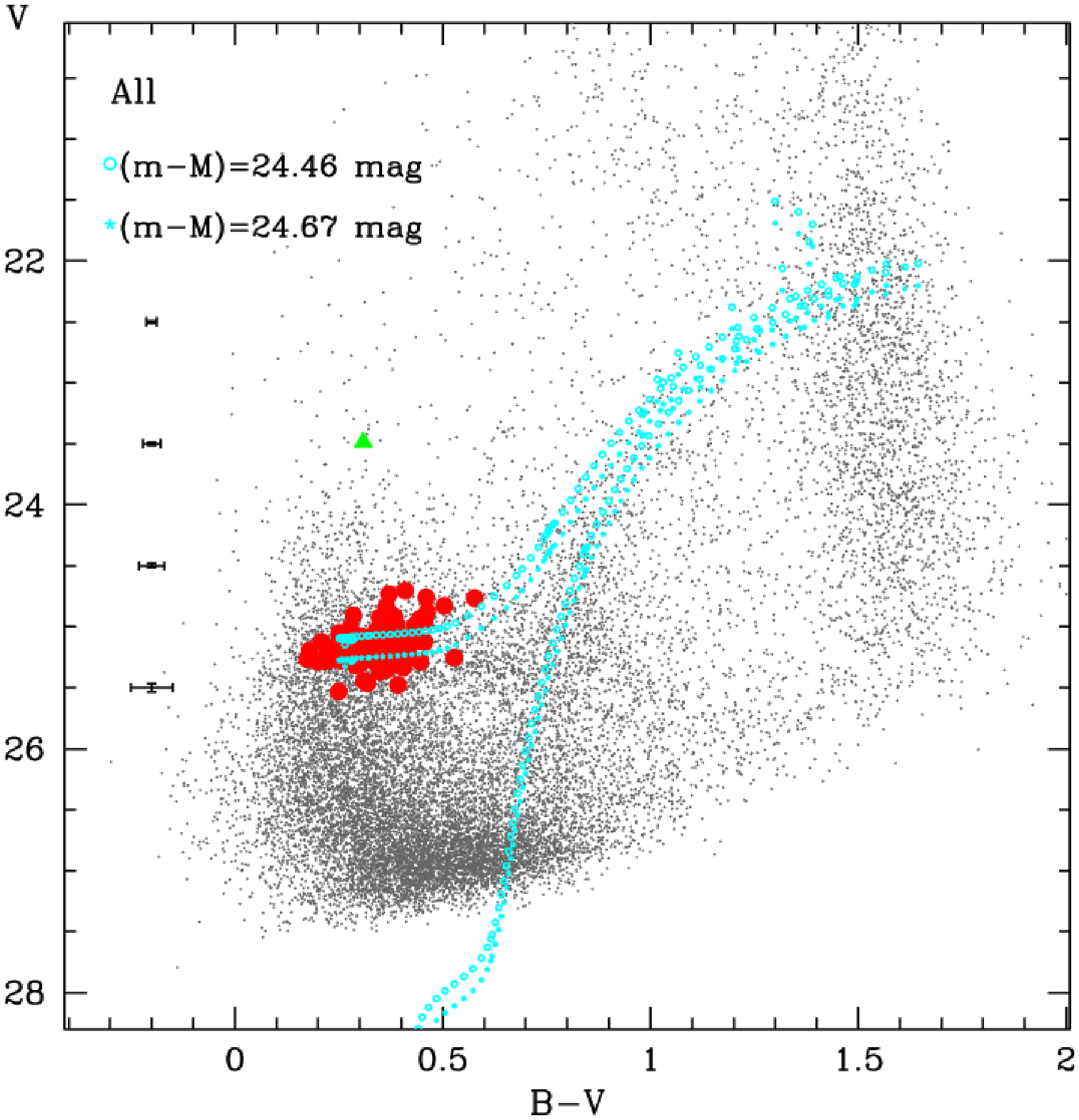}
\caption[]{$Left$: Same as Figure~\ref{fig:cmd} with superimposed the 13 Gyr isochrones from Marigo et al. (2017) 
with $[Fe/H]=-1.8$ dex (cyan  line) and $[Fe/H]=-1.5$ dex (blue line).
$Right$: Same as in the left panel, but with the same isochrone of 13 Gyr and $[Fe/H]=-1.8$  corrected for 
distance moduli of (m-M)=24.46 mag  (upper  line)  and (m-M)=24.67 mag (lower line), respectively. 
}
\label{fig:diffiso}
\end{figure*}

The SE sample appears on average bluer and fainter that the best matching models 
with metallicity from $[Fe/H]=-2.3$ to $-1.7$ dex, M=0.716-0.8 M$_\odot$, T${_eff}$=6000-6900 K and  LogL=1.72-1.76 L$_\odot$
Moreover the observed distribution in periods of the SE RRab  sample is shorter
that the one  derived from the latter models. 
Indeed  models more metal rich than $[Fe/H]=-1.8$  are too red to reproduce the color of the RRab stars.
The observations are only matched if we correct the models for  the distance modulus and reddening found for the SE1 sample 
[(m-M)$_0$=$24.67\pm0.26$ mag, $\langle E(B-V) \rangle=0.03\pm0.05$ mag].
 We  concluded that the scenario for which the RR Lyrae stars have same metallicities, but different distances  is
 best matched by the models
 as shown in the right panel of Figure~\ref{fig:diffiso}. In this case both the RR Lyrae stars 
and the RGB stars are well reproduced by the 13 Gyr isochrones with a $[Fe/H]=-1.8$ corrected for the distance
moduli (m$-$M)$_0$=24.46 mag and (m$-$M)$_0$=24.67 mag, that represent the extreme values reported in Table~\ref{t:sele}.
The twofold RR Lyrae  population can thus not be explained  by the presence of two generations of
stars with different metallicity at the same distance.  The simple old stellar generation
placed at different distances has to be preferred over the double stellar population (see right panel of Figure~\ref{fig:diffiso}). 
We are maybe sampling two different regions in the M31 halo, one belonging to And~XXVII and the 
other to the NW stream that is in the back/front of And~XXVII (see Section~\ref{nwass}).

\section{Association with the North-West stream}\label{nwass}

The NW stream is a stellar structure  extending up to 120 kpc from the center of M31, that 
in its arc-path intersects And~XXVII. This  stream was first detected by \citet{rich11},   
in the PandAS map of  metal poor  ([Fe/H]$<-1.4$ dex) RGB stars.
The mean surface brightness of the NW stream is 28-29 mag arcsec$^{-2}$ 
and the estimated total mass is $\sim 10^7$ M$_\odot$ \citep{carl2011}.
In the literature  a mass  of M=$8.3^{+2.8}_{-3.9}$ $\times 10^7 M_\odot$
was estimated for  And~XXVII  by  \citet{col13} using the velocity dispersion, 
  under the assumption that the galaxy is in dynamical equilibrium and is spherically symmetric.
  However both assumptions are very uncertain given the clearly disrupted nature of this galaxy.
A very rough mass estimate inferred from the total number of RR Lyrae stars  
would place And~XXVII
in the class of $10^7$ $M_\odot$ dwarf galaxies, like And~I or And~VI \citep[see Table~\ref{tab:rrl} for the number of RR Lyrae stars in these galaxies and the masses in][]{col14}.

Projected along the NW stream are also seven GCs. \citet{vel2014}  measured their  
radial velocity (RV) finding that six of them show a clear trend with the 
projected distance from the M31 center, velocities being more negative as the 
GCs approach the M31 center. Monte Carlo simulations performed by \citet{vel2014} 
showed that these six GCs are physically 
associated  among them (with an average RV  of $-430\pm30$ km s$^{-1}$) 
and with the NW stellar stream.
Given its position along the stream, And~XXVII has a RV  compatible 
with that  found  for the GCs 
assuming a quasi-circular inclined orbit where the direction of rotation goes from west to east for the objects in  the NW stream.
And~XXVII maybe be the satellite galaxy that generated the NW stream. The galaxy is  perhaps completing a full
orbit around M31  and after a near pericenter passage  started losing stars.  However, with the current 
data set we can not definitively prove this hypothesis. More kinematic and photometric data are needed to investigate 
the likely connection between the NW stream and And~XXVII.

\section{COMPARISON WITH OTHER M31 SATELLITES AND CONCLUSIONS}\label{concl}
\begin{figure}[t!]
\centering
\includegraphics[trim=40 15 0 0 clip, width=1.05\linewidth]{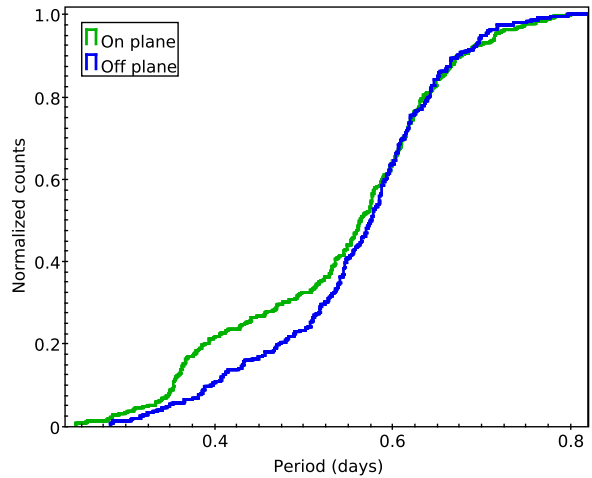}
\caption[]{Cumulative period distribution of the RR Lyrae stars in M31 satellite galaxies
on and off the GPoA.}
\label{fig:cumul}
\end{figure}

Following the procedure in Paper~III we  compared  the pulsation properties of the RR Lyrae populations in the M31 dSph satellites  on and off the GPoA.
Table~\ref{tab:rrl} summarizes the characteristics  of the pulsating variable stars in the M31 dSphs that have been studied so far for variability, 11 in total. 
Six of them are on and five are off the GPoA.%
 We computed the average period of the RRab stars for galaxies on and off the plane separately, finding   $\langle$P$_{\rm ab}\rangle$=0.60$\pm0.07$ d and 
$\langle$P$_{\rm ab}\rangle$=$0.60\pm0.06$ d. The fraction of RRc to total number of 
RR Lyrae stars is f$_c$=0.27 and f$_c$=0.23  for on and off plane satellites, respectively. Hence, both samples 
are compatible with an Oo-Int classification and show a slight tendency to Oo I type as the galaxies get closer to M31  
(see also right panel of Figure~\ref{fig:bayl}). 
The cumulative period distribution of the RR Lyrae stars in the two samples is shown in Figure~\ref{fig:cumul}.
A two sample Kolmogorov-Smirnov (K-S) test to check whether there are any differences between the two populations 
returns a p-value of p= 0.03209, meaning that the two RR Lyrae populations differ significantly. 
  The two samples differ especially in the short period regime (P$<$ 0.4 d, see Figure~\ref{fig:cumul}). This is mainly due to the large number of short period RR Lyrae
  stars in  And~XXVII (indeed  the p value in Paper III was p=0.36, that is without the RR Lyrae stars of And~XXVII). The fraction of RRc stars in And~XXVII  is oddly high.
A similar result was found by \citet{bro04} in some fields of the M31 halo. 
These authors measured  a fraction  f$_c$=0.46, an average period for RRab of $\langle$P$_{\rm ab}\rangle$=$0.594$ d
and an average period for RRc of  $\langle$P$_{\rm c}\rangle$=$0.316$ d.
Between \citet{bro04} results and  ours in And~XXVII (see  Table~\ref{t:sele}) there are some similarities
like the average period of RRab and the high fraction of RRc. Furthermore, the right panel of Figure~\ref{fig:bayl} shows 
a similar trend of the period-amplitude diagram  of M31 halo and And~XXVII RR Lyrae stars.
These analogies can be translated in three 
possible scenarios: 1)  a number of the
RR Lyrae stars  in the field of And~XXVII  do in fact belong to the M31 halo (the projected distance of And~XXVII from the M31 center is d$_p \sim 60$ kpc)
 and in particular to the NW stream.
 This would explain the spread in distance we observe from the RR Lyrae stars; 
2)   And~XXVII is not a dwarf galaxy but rather an overdensity in the NW stream  \citep[as claimed by][]{mart2016}
that is largely composed by M31 halo stars; 3) And~XXVII is the progenitor of the NW stream and part of the M31 halo. 
The high concentration of RR Lyrae stars and the CMD of And~XXVII SE region,  along with the possible connection with the GCs in the NW stream seem to  support this latter possibility.
 However, as shown in Figure~9 of \citet{iba2014}  the M31 halo in And~XXVII neighborhood is very complex and the stars into the metallicity range $-2.5<[Fe/H]<-1.7$ dex
are placed in streams located all around  And~XXVII. Investigation of the variable stars  and CMD of the area beyond the SE region in the south-east direction,
as well as in different regions of the M31 
halo and NW stream is essential in order to discern among the above different scenarios. 

Concerning the frequency of ACs in the  M31 satellites, with the discovery of only one AC in And~XXVII 
we confirm the finding of Paper III that on-plane satellites seem to host only a few or none ACs. There are  4 ACs in total in the 6 on-plane satellites in Table~\ref{tab:rrl} 
to compare with 21 ACs in the 5 off-plane systems. This would suggests that only off-plane satellites were able to retain enough gas to give rise to  an
intermediate-age  stellar  population and   produce  ACs. In this regard, the recent detection of an irregularly shaped HI cloud close to but offset from the stellar 
body of And~XIX (\citealt{kerp2016}) and of some HI emission displaced by half a degree from And~XXI (Kerp private communication), gives support to the possible presence of a 1-2 Gyr old stellar generation in And~XIX  and And~XXI,  and being both these systems off the GPoA, maybe in the off-plane M31 satellites in general.

\begin{table*}
\caption[]{Properties of the variable stars in the Andromeda satellite galaxies}
\footnotesize
\label{tab:rrl}
\begin{center}
\begin{tabular}{l c c l l c c c}
\hline
\hline
\noalign{\smallskip}
Name    & N (RRab+RRc) &  $\langle$P$_{\rm ab}\rangle$  & f$_c$  & N (AC)&N (AC) confirmed*&  member (GpoA) & Reference\\
\hline
And~I   & 72+26       &    0.57   & 0.26 &    1?    &	0  & yes  &   (1)    \\ 
And~II  &   64+8      &   0.57    & 0.11 &    1     &	0  &no  &   (2)    \\
And~III & 39+12       & 0.66      & 0.23 &    5?    &	2  &yes  &   (1)     \\
And~VI  &   91+20     & 0.59      & 0.18 &    6     &	4  &no  &   (3)	  \\
And~XI  &  10+5       &  0.62     & 0.33 &    0     &   0  &yes  &   (4)	 \\
And~XIII& 12+5        &  0.66     & 0.30 &    0     &	0  &yes   &   (4)     \\
And~XVI &   3+6       &   0.64    & 0.33 &    0      &  0  & no$^1$    &  (5,6) \\
And~XIX & 23+8        &   0.62    & 0.26 &    8     &	8  &no  &   (7)   \\	   
And~XXI &   37+4      &   0.63    & 0.10 &    9     &   9   &no  &   (8)  \\
And~XXV &   46+11     &   0.60    & 0.19 &    2	    &   1   &yes  &   (9) \\
And~XXVII & 58+31     & 0.59      & 0.35 &    1     &   1   & yes &   (10)\\
\hline 
\end{tabular}
\end{center}
* confirmed AC, based  on the Period-Weseneheit relation (see Paper~III)\\
$^1$ offset by 8 Kpc from the Great Plane of Andromeda (GPoA, Ibata et al. 2013)\\
(1) \citet{pri05}; (2) \citet{pri04}; (3) \citet{pri02}; \\
(4) \citet{yang12}; (5) Mercurio et al. (2017, in preparation);(6) \citet{mon2016};\\
(7) \citet{cus2013}; (8) \citet{cus2015};(9) \citet{cus2016}; (10) this work
\normalsize
\end{table*}

\acknowledgments

We warmly thank P. Montegriffo for the development and maintenance of the GRATIS software.
Financial support for this research was provided by  PRIN INAF 2010 (PI: G. Clementini) and by Premiale LBT 2013. 
The LBT is an international collaboration among institutions in the United States, Italy, and Germany. LBT Corporation partners 
are The University of Arizona on behalf of the Arizona university system; Istituto Nazionale di Astrofisica, Italy; 
LBT Beteiligungsgesellschaft, Germany, representing the Max-Planck Society, the Astrophysical Institute Potsdam, 
and Heidelberg University; The Ohio State University; and The Research Corporation, on behalf of The University of Notre Dame, University of Minnesota, 
and University of Virginia. We acknowledge the support from the LBT-Italian Coordination Facility for the execution of observations, 
data distribution, and reduction.
Facility: LBT

\end{document}